\documentclass[aps,pre,twocolumn,groupedaddress]{revtex4}
\usepackage{amsmath, graphicx}

\begin{document}

\title{A statistical mechanics framework for static granular matter}

\author{Silke Henkes}
\altaffiliation{Present address: Instituut-Lorentz, Universiteit Leiden, Postbus 9506, 2300 RA Leiden, The Netherlands}
\author{Bulbul Chakraborty}
\affiliation{Martin Fisher School of Physics, Brandeis University, Waltham, MA 02454-9110 USA}

\date{\today}

% need to introduce the ``angoricity'' or whatnot?
\begin{abstract}
The physical properties of granular materials have been extensively studied in recent years. So far, however, there exists no theoretical framework which can explain the observations in a unified manner beyond the phenomenological jamming diagram \cite{Liu}. This work focuses on the case of static granular matter, where we have constructed a statistical ensemble \cite{prl07} which mirrors equilibrium statistical mechanics. This ensemble, which is based on the conservation properties of the stress tensor, is distinct from the original Edwards ensemble and applies to packings of deformable grains. We combine it with a field theoretical analysis of the packings, where the field is the Airy stress function derived from the force and torque balance conditions. In this framework, Point J characterized by a diverging stiffness of the pressure fluctuations. Separately, we present a phenomenological mean-field theory of the jamming transition, which incorporates the mean contact number as a variable. We link both approaches in the context of the marginal rigidity picture proposed by \cite{Wyart,Ellenbroek}.
\end{abstract}

% still need the pacs numbers
%\keywords{}

\maketitle

\section{Introduction \label{chap:introduction}}

\subsection{Granular media}
Granular materials are ubiquitious in everyday life, as sand, salt or rice; or as the breakfast cereal in your bowl. They are also of fundamental importance in industrial processing, from the transport of coal to drug manufacturing. The shape of the surface of the Earth is largely determined by the erosion of rock to sand and dust and the patterns they form subsequently. If we want to truly understand the world we live in, understanding granular materials must be part of our efforts.
Yet, our grasp of the behavior of granular matter is limited \cite{deGennes_review,Jaeger_Nagel_review,khakhar_review,Kadanoff_review}. A consensus about what the important questions in the field are, and on suitable approaches to answer them, has only started to emerge in the last few years \cite{Liu, Bulbul_Bob}.

The chief obstacle that hinders our understanding is that granular matter is fundamentally out of thermal equilibrium, since the typical interaction energies of the particles are many orders of magnitude larger than $k_{b}T$. Therefore, granular materials are effectively at $T=0$ from a conventional point of view. This means that in equilibrium, the system does not explore the space of states available to it, thus leading to a breakdown of ergodicity. It also implies that the Boltzmann ensemble is irrelevant, and we are left without the framework of equilibrium statistical mechanics to guide our understanding. 

Since granular systems do not equilibrate spontaneusly, the method of preparation of a granular packing matters, i.e. granular materials are history-dependent. A steady-state can only be achieved through highly controlled experimental conditions like repeated tapping \cite{Jaeger_Nagel_compaction} or compression \cite{Makse_compaction}. This complicates the study of granular materials since it is far from clear if a universal framwork is even possible.

\subsection{Towards a non-equilibrium theory}
Despite the obstacles described in the previous section, some progress has been made towards a non-equilibrium theory of granular materials. Most theoretical work has attempted to extend concepts like temperature or entropy from the thermodynamic context to the granular context.

It has been recognized that granular packings and granular systems under shear or other forms of agitation share several important properties of glassy systems. Like granular materials, glassy materials have fallen out of equilibrium in the sense that thermal fluctuations are too small to allow the system to undergo rearrangements on short time scales. Some model glass formers, like Lennard-Jones fluids, are structurally similar to granular packings. A common concept that has emerged for both systems is the idea of a rugged potential energy landscape with a myriad of energy minima. For glassy systems, these are termed the inherent structures \cite{Stillinger} in which the system spends a long time before escaping over an energy barrier to a different one via an activated process. For granular systems, the potential energy minima correspond to packings in mechanical equilibrium, and the dynamics are provided by tapping or slow shearing. For situations where the deformation is gradual enough for the system to move by going through a sequence of separate rearrangements (the quasistatic shear limit), Bouchaud's trap model for glasses \cite{Bouchaud} has been adapted to include shearing as the SGR (Soft Glassy Rheology) model by Sollich \cite{Sollich}.

The liquid to solid phase transition in granular media is also known as the jamming transition. It is controlled by parameters including an external driving force like shaking or shearing, but that do not include temperature. This transition shares some of the properties of the glass transition. In the glass transition the system changes from a liquid to a disordered solid without an obvious change in microscropic structure, unlike for the conventional liquid-solid transition where the system goes from a state of high symmetry (the liquid phase) to a state of lower, broken symmetry (the crystalline solid) \cite{Chaikin_Lubensky}.

These similarities have prompted Liu and Nagel to propose \cite{Liu} a unified phase diagram for glassy and granular materials, as well as related materials like foams, bubbles and colloids. Glassy materials are in the temperature - density plane of the diagram. The transition between a granular solid and a granular liquid with increasing shear is in the density-shear plane. The packing fraction axis shared by both types of materials is characterized by granular packings in mechanical equilibrium, or alternatively glassy systems at zero temperature. 

The properties of the phase transition depend on how the phase boundary is approached. Along the packing fraction axis, reducing the density of an isotropic jammed assembly of grains leads to a point at which both the bulk modulus and the shear modulus, i.e. the resistance of the system to uniform compression and shear, vanish \cite{Behringer_Nature,epitome}. The transition point has been termed Point J by \cite{Liu}, and it has some of the features of a critical point, though it not obvious that Point J has any bearing on the glass transition \cite{Witten-Berthier}.

O'Hern et al. \cite{epitome} have found power-law scalings and apparently universal properties when J is approached from the high-density limit, and scaling as well as a divergent length scale have also been observed by Drocco et al. \cite{Reichhardt} and Teitel et al. \cite{Teitel} when J is approached from the low-density limit. Experimentally, the microscopic properties of static packings, such as the force and contact number distributions, as Point J is aprroached along the packing fraction axis have been studied by Majmudar et al. \cite{Trush}.

For spherical particles, Point J can be linked to the underlying microscopic properties of the granular packings. For spherical granular materials with purely repulsive, short-range interactions, Point J is identical to the \emph{isostatic point}, the point where the number of degrees of freedom is equal to the number of constraints imposed by the intergranular forces \cite{epitome}. In the frictionless case, the jamming thresholds of different configurations converge with increasing system size to a single packing fraction \cite{epitome} which is identical to \emph{random close packing} (RCP), the highest packing fraction that can be attained by random packings of hard grains \cite{Berryman_RCP}. For three dimensional packings of spheres, $\phi_{J}=0.63$, while for two dimensional packings of disks it is given by $\phi_{J}=0.84$.

A majority of experimental studies of the jamming transition probe the transition as a function of the shear rate, where the phase boundary is crossed at finite positive pressure. In this case, only the shear modulus vanishes at the transition and the nature of the transition is different from the compression case. The signature of this transition is intermittent jamming and unjamming with large-scale stress fluctuations, and pronounced dilatancy effects\cite{Bulbul_Bob}. A different type of jamming occurs in systems where there is a sustained motion of the grains, and kinetic energy plays a role. These transitions are characterized by the appearence of large scale spatio-temporal fluctuations in the particle motion \cite{Menon,Ally}. These fluctuations have been termed \emph{dynamical heterogeneities}, after the very similar behavior which is observed in supercooled liquids \cite{Berthier_glass}. Further analysis using tools developed for the glass transition, like the self-intermediate scattering function, has revealed a growing length scale as the jamming transition is approached \cite{Ally_07,Dauchot_dynhet,Durian}.

\subsection{Granular statistical mechanics}
Going further in the analogy to thermodynamics and ordinary statistical mechanics, several attempts have been made to build a framework equivalent to equilibrium statistical mechanics for granular materials.

The first attempt was by Edwards \cite{Edwards_Oakeshott} who replaced the energy conservation law for thermal systems with a volume conservation law for granular packings. Edwards makes the microcanonical assumption that all states at the same free volume $V$ are accessed with the same probability. For a subsystem with volume $v$ of the full packing, one then obtains a canonical distribution for the probability to access the state
\begin{equation} P_{\nu} = \exp(-v_{\nu}/\chi(V)); \label{Edwards} \end{equation}
where the Lagrange multiplier $\chi$, or the \emph{compactivity}, plays the role of temperature. $\chi$ is the best known example of a granular temperature for static packings (see \cite{Edwards_Grinev} for a review).

The validity of the Edwards ensemble has been investigated for experimental and simulated grain packings. Some studies have been direct investigations of the free volume distribution in packings \cite{Makse_vol_entropy, Blumenfeld_Edwards_3d, Dauchot_volume, Aste_diMatteo}. 
% need to say about issues and what found
A lot of the tests have been indirect, generally through the \emph{fluctuation-dissipation theorem} \cite{Chaikin_Lubensky} (see \cite{Makse_tracer} and \cite{Makse_Kurchan_Nature}).

Other out-of-equilibrium temperatures , mostly based on the fluctuation-dissipation theorem, have been proposed in the context of dynamical granular media. One study \cite{effective_temperatures} measured several of them in a simulated system of a two-dimensional sheared foam, and found that the different temperatures agree with each other, showing that effective temperatures are a useful concept for driven granular systems.

In recent years, it has become more common to apply the Edwards equiprobability hypothesis to force configurations on static granular packings (see e.g. \cite{Socolar}, \cite{Metzger} and \cite{vanHecke}). An equivalent approach to the case of the volume can be taken, and a force-based canonical ensemble is then introduced. The resulting granular temperature has been dubbed the \emph{angoricity} by Edwards \cite{Edwards_Blumenfeld}.

\subsection{This work}
Here we derive the conservation law responsible for the force ensemble from first principles, and show the conditions under which the canonical ensemble is valid (see also our previous work \cite{prl07}).
We construct and examine a statistical mechanics-like framework in which Point J and the properties of jammed systems can be studied. We restrict our attention to packings on the density-shear plane of the jamming phase diagram, where we find a conservation law upon which we base an ensemble in which stress plays the role of energy. We then use the basis laid out by the ensemble to develop a field theoretical model for jammed packings, which predicts a diverging stiffness at the jamming transition. Finally, we discuss an empirical mean-field theory derived from the analysis of simulated packings.

% \paragraph{Stress ensemble}
Section 2 derives a conservation law for the force-moment tensor, i.e. the volume integral of the stress tensor, from the boundary property shown in \cite{BB,prl07}. This forms the basis for the derivation of the stress ensemble, where the force moment tensor plays the role of energy, from first principles. We derive a tensorial analog of temperature,  the angoricity, which is related to stress fluctuations rather than energy.  The concept of angoricity has been introduced earlier.  To avoid any confusion, we precisely define our usage of the term in this paper. We also discuss the case of an isotropically compressed system where we can simplify the ensemble to a scalar stress variable and a scalar temperature. The stress ensemble has been numerically tested in \cite{prl07}, and we provide a summary of the results. Finally, we discuss an exact model of the jamming transition compatible with the universal equation of state found in the simulation.

% \paragraph{Field theory}
Section 3 builds a field theoretical model of granular packings based on the scalar stress function appropriate for isotropic compression. Based on symmetry arguments and on comparison to simulation results, we find that the packings can be described by a laplacian field theory with a stiffness that depends on the compression. The stiffness diverges at the jamming transition, so that the entropy defined by the number of packings available vanishes. Finally, we discuss shear fluctuations in the context of the field theory.

% \paragraph{Mean field theory of the jamming transition}
Section 4 combines the equation of state found empirically with the form of the density of states. It incorporates another empirical relation linking stress and contact numbers, which has also been derived from a stability argument, to define an effective free energy. We study the predictions of the effective free energy for the jamming transistion, and find that the theory admits divergent fluctuations at Point J; but only if formulated in terms of a variable $u$ measuring the deviations from the stability line. We conclude by showing that the simulated system indeed shows a divergence in the fluctuations of $u$ as Point J is approached.

%*************************************************************************************************************
\section{The stress ensemble \label{chap:ensemble}}
In this section, we derive from first principles an ensemble based on the conservation of the total force moment tensor for static packings of granular materials under given boundary conditions. This stress ensemble provides the formal basis of the force network ensemble used in \cite{vanHecke,Socolar}. We have already presented a derivation of the ensemble in \cite{prl07}, however the derivation below has been improved by taking into account the general formalism for nonequilibrium temperatures presented by Bertin et al. \cite{DauchotPRE}, and it has been generalized to the full force moment tensor $\hat{\Sigma}$.

\subsection{Conserved quantities in stable grain packings} 

The total stress tensor of a granular packing with volume $V$ and only contact forces can be written as \cite{BB}
\begin{equation} \hat{\sigma} = \frac{1}{V} \sum_{\langle ij \rangle} \vec{r}_{ij}\vec{F}_{ij} \equiv \frac{\hat{\Sigma}}{V}, \end{equation}
where the sum runs over all the contacts in $V$, the $\vec{F}_{ij}$ are the contact forces between grains $i$ and $j$ and the $\vec{r}_{ij}$ are the distances between grain centers and contact points.
The extensive quantity $\hat{\Sigma}$, the total force moment tensor, depends only on the boundary conditions of the packing which determine $\hat{\sigma}$ and the total volume $V$. This fundamental result is ultimately a consequence of force balance, and it can be demonstrated in a more formal way as shown below:

Ball and Blumenfeld \cite{BB} have shown that local force balance in a two-dimensional granular packing can be incorporated through the definition of a vector height field field, $\vec{h}_{\mu}$, on the loops $\mu$ dual to the contact network.
With this definition, the stress tensor associated with an area $A$ can be written as a boundary term that only involves the boundary vectors $\vec{d}_{\mu_{b}}$ and the boundary height field $\vec{h}_{b}$, see also Figure 1 in \cite{prl07}. In three dimensions, the same result can be obtained by a continuum approach involving a tensorial equivalent of the height field \cite{Gurtin}.

% Need to see if we need to inlude the DataHeight figure again ......
The global force moment tensor is then (the first expression is for two-dimensional systems, while the second expression is for three dimensional sytems)
\begin{equation} \hat{\Sigma} =_{2d} \sum_{\text{bound.}} \vec{d}_{\mu_{b}} \vec{h}_{\mu_{b}}  =_{3d} \int_{\partial V} \hat{W} \times \vec{n} \: dA. \label{global_force_moment}
\end{equation}
which is the global equivalent to writing the stress tensor as the curl of the height field $\hat{\sigma} = \vec{\nabla} \times \vec{h}$ ($\hat{\sigma} = \vec{\nabla} \times \hat{W}$ in three dimensions).

The fact that $\hat{\Sigma}$ can be written as a boundary integral implies a conservation principle for force-balanced, static grain configurations:
each rearrangement of the grains within a region of packing, so long as it does not affect the boundary of the region, will not change the total force moment $\hat{\Sigma}$ of the region.

The force moment tensor is an extensive quantity, it is additive if we neglect boundary effects, as is the case for energy in thermodynamics, and it is conserved for any dynamics which keeps the boundary conditions of a granular packing intact. We now use $\hat{\Sigma}$ to define a canonical ensemble, equivalent to the Boltzmann ensemble, where it plays the role of the energy (incidentally, $\hat{\Sigma}$ also has the dimensions of an energy).

An example of rearrangements which conserve the total force moment tensor are the wheel moves introduced by Tighe et al. on the triangular lattice \cite{Socolar}.

\subsection{Derivation of the canonical stress ensemble}

\paragraph*{Counting the number of states}
The first step towards constructing an ensemble equivalent to the Boltzmann ensemble is to define a density of states. The total force moment tensor $\hat{\Sigma}$ of a compact area $A_{N}$ with $N$ grains is not affected by internal rearrangements. Hence we can partition the phase space of all the possible force- and torque-balanced packings on $A_{N}$ (the blocked states) into sectors $M_{\hat{\Sigma}}$  with the same total force moment tensor $\hat{\Sigma}$. We then count the number of states in each sector $M_{\hat{\Sigma}}$, $\rho_{N}(\hat{\Sigma})$. This number includes all the possible geometric configurations of the grains and will thus always be greater than $1$, even at the isostatic point. 

In a thermal system, $\rho_{N}(E)$, the total number of configurations at a given energy, would form the basis of the microcanonical ensemble. A fundamental assumption underlying thermodynamics and equilibrium statistical mechanics is that the system is ergodic such that all states with a given energy are visited equally. The probability of finding a state is then just a flat measure in state space, i.e $p_{\nu}|_{E} = 1/\rho_{N}(E)$, and we can then identify the density of states as $\Omega_{N}(E) = \rho_{N}(E)$. 

In a granular system, the dynamical processes which move the system from one state to the next are highly varied and to a certain degree arbitrary. States within a sector can (but may not) be connected by local dynamics; but states in different sectors are inaccessible through purely local dynamics. We can think of the phase space of blocked states through a \emph{landscape}, similar to the energy landscape often invoked for glassy materials. However, it is the chosen dynamics, not thermal fluctuations, that move the system from one blocked state to another. The dynamics are in general non-ergodic, and in the extreme case of a purely static system, we recover the $T=0$-limit of the glass.

Detailed balance is violated, and hence the system is not in thermal equilibrium and the states are not necessarily accessed with equal probability. Therefore, the density of states for $M_{\hat{\Sigma}}$ is not only dependent on the number of states $\rho_{N}(\hat{\Sigma})$ at that $\hat{\Sigma}$, but also on the \emph{frequency} $\beta_{\nu}^{(dyn)}$ with which each state is accessed by the dynamics chosen to create the packing. The quantity equivalent to the density of states is then given by:

\begin{equation} \Omega(\hat{\Sigma}) = \sum_{\nu \in M_{\hat{\Sigma}}} \beta_{\nu}^{\text{dyn}} \label{DOS_definition}\end{equation}
\\

\paragraph*{Preconditions}
We develop the ensemble along the same lines as for a thermal system \cite{chandlerbook}. Bertin et al. \cite{DauchotPRE} have shown that intensive quantities, which can be interpreted as nonequilibrium temperatures, can be defined in a system in steady-state, so long as there is an additive conserved quantity in the system. The conserved quantity plays the role of energy, and leads then to a temperature variable conjuguate to it. We follow their method of derivation below.

For mechanically stable granular packings, $\hat{\Sigma}$ is the additive quantity that replaces the energy. The following results are dynamics-dependent, and stay valid for \emph{any} dynamics which satisfies equation \ref{beta_fac}, but only as long as the \emph{same} dynamics are used in the preparation of any systems we wish to compare (e.g. in steady-state). Whether different types of dynamics can give rise to the same density of states, and if so, which classes do, is a question which we address at the end of this section.

Consider a system $S$ which contains $N$ grains, with total force moment tensor $\hat{\Sigma}$. We partition the system into two compact subsystems $S_{1}$ and $S_{2}$ with grain numbers $N_{1}$ and $N_{2}$, and total force moment tensors $\hat{\Sigma}_{1}$ and $\hat{\Sigma}_{2}$, respectively. Since $\hat{\Sigma}$ is additive, we always have $\hat{\Sigma} = \hat{\Sigma}_{1} +\hat{\Sigma}_{2}$. 
\\

\paragraph*{The microcanonical ensemble}
If we fix the total force moment tensor $\hat{\Sigma}$ in $S$, the conditional probability of finding a force moment tensor $\hat{\Sigma}_{1}$ in subsystem $S_{1}$ is defined by $P(\hat{\Sigma}_{1}) = P(\hat{\Sigma}_{1}|\hat{\Sigma}) P(\hat{\Sigma})$. Then we can use the definition of the density of states equation \ref{DOS_definition} to write the conditional probability as
\begin{equation*}  P(\hat{\Sigma}_{1}|\hat{\Sigma}) = \Omega_{N}(\hat{\Sigma})^{-1}\!\!\! \sum_{\nu \in M_{\hat{\Sigma}}} \!\beta_{\nu}^{\text{dyn}} \delta (\hat{\Sigma}_{\nu_{1}} \!- \hat{\Sigma}_{1})\delta (\hat{\Sigma}_{\nu_{2}} \!- (\hat{\Sigma}-\hat{\Sigma}_{1})\!). \end{equation*}

If the frequency with which the subsystems are accessed factorizes, i.e. if 
\begin{equation} \beta_{\nu}^{(dyn)}=\beta_{\nu_{1}}^{(dyn)}\beta_{\nu_{2}}^{(dyn)}, \label{beta_fac} \end{equation} the conditional probability becomes
\begin{equation} P(\hat{\Sigma}_{1}|\hat{\Sigma}) = \frac{\Omega_{N_{1}}(\hat{\Sigma}_{1}) \Omega_{N_{2}}(\hat{\Sigma}-\hat{\Sigma}_{1})}{\Omega_{N}(\hat{\Sigma})} \end{equation}
as a function of the densities of states of the subsystems.
Equation \ref{beta_fac} is the central assumption in the derivation of the stress-ensemble, and it is conceptually equivalent to requiring that the dynamics choose state $\nu_{1}$ of $S_{1}$ independently of state $\nu_{2}$ of $S_{2}$. Since the subsystems interact through their shared boundary only, we expect the correction to equation \ref{beta_fac} to scale as $\mathcal{O}(1/\sqrt{N_{1}})$, where $N_{1}$ is the number of grains in $S_{1}$. However, If the system is correlated over length scales $\xi \geq \sqrt{N_{1}}$, we expect equation \ref{beta_fac} to break down as well. We discuss the validity of this assumption, and how we have tested it, at the end of this section.

The most probable value $\hat{\Sigma}_{1}^{*}$ at a given $\hat{\Sigma}$ is found by setting the derivative of the conditional probability with respect to $\hat{\Sigma}_{1}$ to zero. Since $\hat{\Sigma}_{1}$ is a tensor, this needs to be done for each component separately. We use the logarithmic derivative to simplify the calculation,
\begin{equation*} 
0 = \frac{\partial \ln \Omega_{N_{1}}(\hat{\Sigma}_{1})}{\partial \hat{\Sigma}_{1}^{ij}}|_{\hat{\Sigma}_{1}^{ij*}}+\frac{\partial\ln \Omega_{N_{2}}(\hat{\Sigma}-\hat{\Sigma}_{1})}{\partial \hat{\Sigma}_{1}^{ij}}|_{\hat{\Sigma}_{1}^{ij*}} 
\end{equation*}
where we have replaced the derivative in the $\Omega_{N_{2}}$-term by $\partial_{\hat{\Sigma}_{2}^{ij}} = -\partial_{\hat{\Sigma}_{1}^{ij}}$. The first and the second term are then opposites of each other, and we define the microcanonical equivalent of the inverse temperature $\alpha_{ij}$ by
\begin{equation} \hat{\alpha}_{ij} =\frac{\partial \ln \Omega_{N_{1}}(\hat{\Sigma}_{1})}{\partial \hat{\Sigma}_{1}^{ij}}|_{\hat{\Sigma}_{1}^{ij*}}=\frac{\partial\ln \Omega_{N_{2}}(\hat{\Sigma}_{2})}{\partial \hat{\Sigma}_{2}^{ij}}|_{\hat{\Sigma}-\hat{\Sigma}_{1}^{ij*}}\!. \label{alpha_microcan} \end{equation}
We will show below that $\hat{\alpha}_{ij}$ acts indeed like an inverse temperature, in that it is independent of the partitioning of $S$.
\\

\paragraph*{The canonical ensemble}
To define the canonical ensemble, we consider the same system $S$, but we divide it now into one very small partition $S_{m}$, with $m<<N$, and the remaining $S_{N-m}$ (note that we still need $m>>1$ for the factorization of the density of states to hold). We focus our attention now on the small partition where the total stress can fluctuate while the full system $S$ with fixed $\hat{\Sigma}$ acts as a reservoir, similar to the thermal case. We still have
\begin{equation*} P(\hat{\Sigma}_{m}|\hat{\Sigma}) = \frac{\Omega_{m}(\hat{\Sigma}_{m}) \Omega_{N-m}(\hat{\Sigma}-\hat{\Sigma}_{m})}{\Omega_{N}(\hat{\Sigma})}. \end{equation*}
Taking the logarithm of both sides, we can expand $\Omega_{N-m}(\hat{\Sigma}-\hat{\Sigma}_{m})$ to first order in $\hat{\Sigma}_{m}$ and find that
\begin{equation} \ln P(\hat{\Sigma}_{m}|\hat{\Sigma}) = \ln \Omega_{m}(\hat{\Sigma}_{m}) - \sum_{l,k =1}^{d} \frac{\partial \ln \Omega_{N}(\hat{\Sigma})}{\partial \hat{\Sigma}^{ij}} \hat{\Sigma}_{m}^{ij} \label{taylor_DOS} \end{equation}
We define the canonical inverse temperature tensor by
\begin{equation} \hat{\alpha}_{ij} = \frac{\partial \ln \Omega_{N}(\hat{\Sigma})}{\partial \hat{\Sigma}^{ij}} \label{alpha} \end{equation}
and then the sum in equation \ref{taylor_DOS} becomes the total contraction $\hat{\alpha}_{ij}\hat{\Sigma}_{m}^{ij} = \text{Tr}(\hat{\alpha}^{T}\hat{\Sigma}_{m})$. The order of the indices is irrelevant since $\hat{\Sigma}$ is a symmetric tensor, and so is $\hat{\alpha}$ through its definition.
The canonical probability distribution for the total force moment tensor $\hat{\Sigma}_{m}$ is then 
\begin{equation} P^{\text{can}}(\hat{\Sigma}_{m}) = P(\hat{\Sigma}_{m}|\hat{\Sigma}) = \Omega_{m}(\hat{\Sigma}_{m}) \frac{e^{-\text{Tr}(\hat{\alpha}\hat{\Sigma}_{m})}}{ Z(\hat{\alpha})} \label{can_distribution_tensor}\end{equation}
The canonical partition function 
\begin{equation}
Z_{m}(\hat{\alpha}) = \prod_{l, k>l}\int d\hat{\Sigma}_{m}^{lk} P^{\text{can}}(\hat{\Sigma}_{m}) 
\end{equation}
 acts as a normalization.

The angoricity defined by Edwards\cite{Edwards_Blumenfeld} is related to the {\it inverse} of $\hat{\alpha}$.   In its original definition, the angoricity is defined as the derivative of the {\it stress tensor} with respect to the entropy.  From equation \ref{alpha}, $\hat{\alpha}$ is the derivative of the entropy with respect to the force moment tensor.  In the interest of controlling the proliferation  of terms associated with temperature-like quantities for granular materials,  we refer to the inverse of $\hat{ \alpha}$ as the angoricity. The modified Boltzmann factor for the granular system is then $\exp(-\text{Tr}(\hat{T}^{-1} \hat{\Sigma}_{m}))$.
Finally we show that the inverse granular temperature in the canonical ensemble and in the microcanonical ensemble are equal. If we repeat the derivation of (\ref{alpha_microcan}) using the form (\ref{can_distribution_tensor}) of the canonical distribution, then we obtain
\begin{equation*} \frac{\partial \ln P(\hat{\Sigma}_{1}|\hat{\Sigma})}{\partial \hat{\Sigma}_{1}^{ij}} |_{\hat{\Sigma}_{1}^{ij *}} =  \frac{\partial \ln \Omega_{N_{1}}(\hat{\Sigma}_{1})}{\partial \hat{\Sigma}_{1}^{ij}}|_{\hat{\Sigma}_{1}^{ij*}} - \hat{\alpha}(\hat{\Sigma})_{ij} = 0 \end{equation*}
So the microcanonical inverse temperature equals the canonical inverse temperature, and is independent of the partitioning of the system.
\\

\paragraph*{Special case of an isotropic system}
A lot of experimental and simulation effort has been devoted to systems under hydrostatic pressure \cite{epitome,Behringer_Nature,Dinsmore} caused by fixing the volume of the system in the absence of shear.
Let $\Gamma = \text{Tr}(\hat{\Sigma})$ be the trace of the force moment tensor. The extensive variable $\Gamma$ is related to the intensive pressure by $\Gamma= pA$, where $A$ is the area of the system. This makes $\Gamma$ the internal virial of the system \cite{internalvirial-ref}. Then we can write the force moment tensor for an isotropic system in the absence of shear as $\hat{\Sigma} =  \frac{\Gamma}{2} \hat{I}$ \cite{Landau}.
It is natural to simplify the formalism to a single scalar variable $\Gamma$. Since the trace is additive, $\Gamma$ is still a conserved, additive variable and the microcanonical and canonical ensemble derivations are the same as above. The key results are the definition of $\alpha$
\begin{equation} \alpha = \frac{\partial \ln \widetilde{\Omega}_{N}(\Gamma)}{\partial \Gamma} \label{alpha_scalar}\end{equation}
and the form of the canonical distribution
\begin{equation} P^{\text{can}}(\Gamma_{m}) = P(\Gamma_{m}|\Gamma) = \widetilde{\Omega}_{m}(\Gamma_{m}) \frac{e^{-\alpha\Gamma_{m}}}{ Z(\alpha)} \label{can_distribution}\end{equation}
with the partition function
\begin{equation}
Z(\alpha) = \int d\Gamma P^{\text{can}}(\Gamma_{m}). \label{can_scalar_partition}
\end{equation}
We now show that $\alpha$ is related to the tensorial inverse temperature $\hat{\alpha}$ by $\hat{\alpha} =\alpha \hat{I}$:
If we consider equation \ref{alpha} for an isotropic system, the density of states $\Omega_{N}(\hat{\Sigma})$ must be symmetric under $\hat{\Sigma}^{12} \rightarrow -\hat{\Sigma}^{12}$ since no direction of shear is preferable to another. Therefore, $\Omega_{N}(\hat{\Sigma})$ has an extremum at $\hat{\Sigma}^{12}=0$, so that the logarithmic derivative with respect to $\hat{\Sigma}^{12}$ vanishes for the shear-free system and we obtain $\alpha_{12} = 0$ (and by extension $\alpha_{21}=0$ since $\hat{\alpha}$ is symmetric). Likewise, the density of states must be invariant under rotations, so that the derivatives with respect to $\hat{\Sigma}^{11}$ and $\hat{\Sigma}^{22}$ are the same: $\alpha_{11}=\alpha_{22} =\alpha$. Then the Boltzmann factor $\exp(-\text{Tr}(\hat{\alpha}\hat{\Sigma}))$ becomes $\exp(-\alpha \Gamma)$. The density of states for $\Gamma_{m}$, $\widetilde{\Omega}_{m}(\Gamma_{m})$, can be related to $\Omega_{m}(\hat{\Sigma}_{m})$ by using $P^{\text{can}}(\Gamma_{m}) = \sum_{\hat{\Sigma}_{m}} P^{\text{can}}(\hat{\Sigma}_{m}) \delta(\Gamma_{m}-\text{Tr}(\hat{\Sigma}_{m}))$ and we obtain
\begin{equation} \widetilde{\Omega}_{m}(\Gamma_{m}) = \prod_{l,k>l}\int d\hat{\Sigma}_{m}^{lk}\; \Omega_{m}(\hat{\Sigma}_{m}) \delta(\Gamma_{m}-\text{Tr}(\hat{\Sigma}_{m})) \end{equation}

\subsection{Discussion \label{sec:ensemble_theo_discussion}}
It is important to ask if there are important differences between angoricity and temperature. The most important distinction between a granular system and a thermal system is that the granular system has to be \emph{driven} to change configurations. There is no simple equivalent to the thermal agitation which serves as a temperature bath for equilibrium systems, and which gives a natural value for the temperature. 

A granular system that is slowly sheared, so that it changes configurations based on the imposed strain, seems to come close to a thermal system. The external load resulting from the shearing sets the scale of the granular temperature, and the off-diagonal parts of $1/\alpha$ can then be seen as a measure of the strength of the perturbations that the shear inflicts on the force chains. In this picture, the system stays in a force- and torque-balanced configuration until the load on a force chain becomes too large, upon which the system rearranges itself into another configuration in mechanical equilibrium. 
Over time, the packing visits a large number of configurations in a \emph{stress landscape} analogous to the energy landscape for glassy systems. The dynamics of an equivalent system to the one described here, but in an energy landscape, is the subject of the SGR (Soft Glassy Rheology) theory \cite{Sollich}. Recent work based on a toy model of activated dynamics in a stress landscape\cite{bob_toymodel} has been compared to experiments\cite{Behringer_logshear} as a test of the stress ensemble. The adaptation of the full SGR to the stress ensemble is in preparation \cite{Max}.

All of the above derivations can be performed in an equivalent manner for the volume as the extensive, conserved variable. The question of how the frequencies $\beta^{\text{dyn}}_{\nu}$ with which the different microstates are accessed depends on the experimental or simulation protocol poses itself both for the stress ensemble and the Edwards ensemble.

In the Edwards ensemble paradigm, this question was explored by O'Hern et al. \cite{Corey_notequal} for very small disk packings $N\leq 14$ by enumerating all the states and measuring their frequencies for two different protocols. The same states were found with both methods, but as expected with different frequencies. Thus the microcanonical equiprobability assumption is violated in this case, and there is no reason to expect a different result for the stress ensemble.

We still have to investigate the validity of equation \ref{beta_fac}, the factorization of the density of states. This is a non-trivial assumption, since it breaks down if there are correlations between the subsystems we consider. For a volume-based ensemble derived along the same lines as above, Lechenault et al. have shown experimentally that even for sub-systems of size $m\!>\!100$, there are corrections to equation \ref{beta_fac}.

A method to investigate the stress ensemble is similar to the approach taken for the Edwards ensemble. Where for the Edwards ensemble, the sytem is repeatedly compactified at the same volume \cite{Jaeger_Nagel_compaction}, so that after each compression it enters a new configuration, we can create a new system with the same boundary stresses at each step. We now give a summary of our previous \cite{prl07} tests of the ensemble on simulated packings which are created from a random initial state and then relaxed until they reach mechanical equilibrium.

\subsection{Summary of simulation results}
We have tested the stress ensemble formalism on simulated packings of frictionless disks with either harmonic or hertzian interactions in two dimensions \cite{prl07}, using the algorithm of O'Hern et al. \cite{epitome}. We first rescaled the $\Gamma$-distributions of different configurations to test the form of the canonical distribution equation \ref{can_distribution}. We find that we are able to perform the rescaling for any subsystem larger than $m>3$, and that the equation of state $\alpha(\Gamma)$ we extract close to the unjamming transition$\alpha$ $\langle z \rangle =z_{iso}=4$ also becomes $m$-independent above this value. From this we conclude that the factorization property equation \ref{beta_fac} which underlies equation \ref{can_distribution} is valid for $m>3$.

\begin{figure}[h]
\begin{center}
\includegraphics[width = 0.8\columnwidth,trim = 2mm 0mm 15mm 0mm, clip]{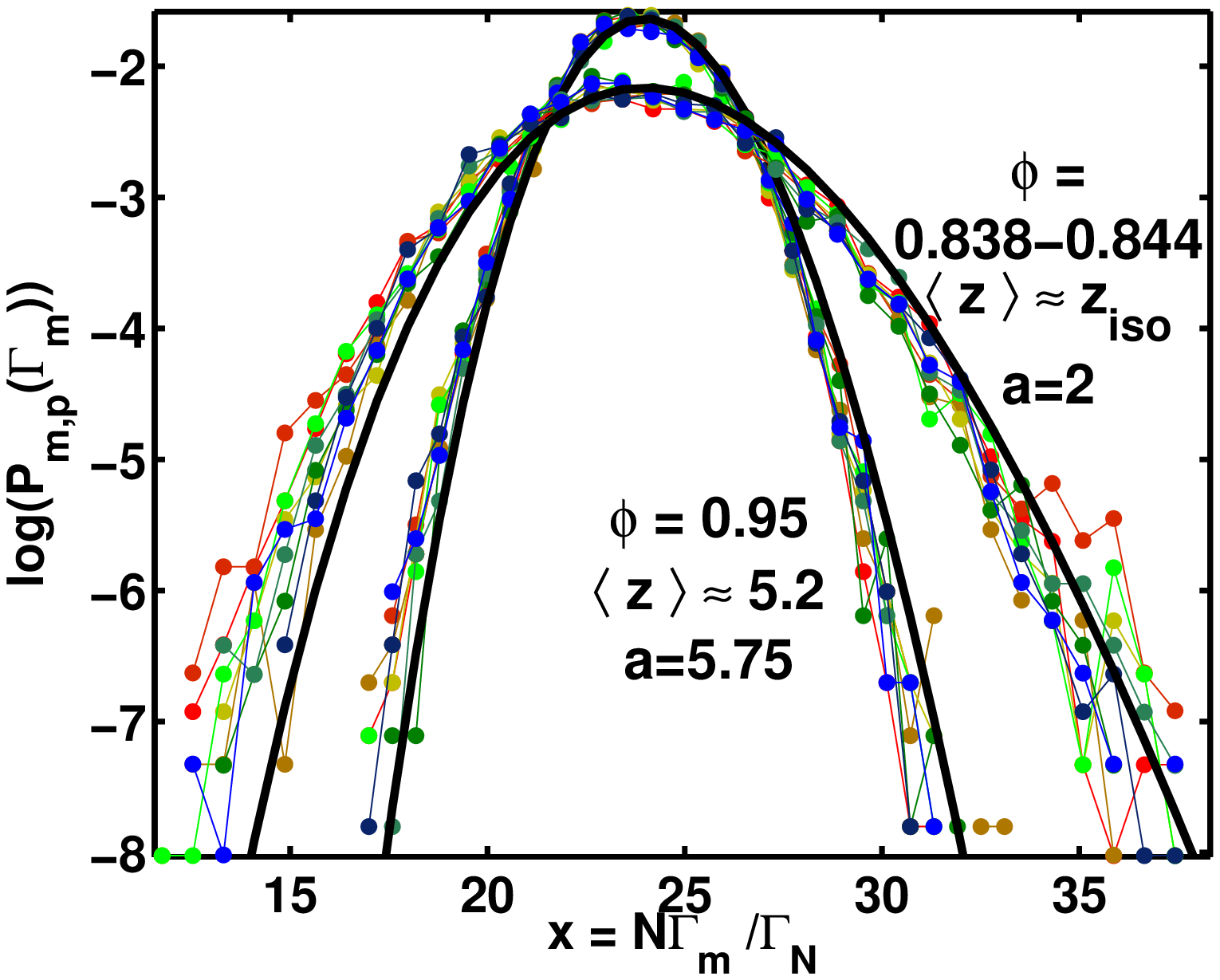}
\includegraphics[width = 0.8\columnwidth, trim = 5mm 0mm 15mm 0mm, clip]{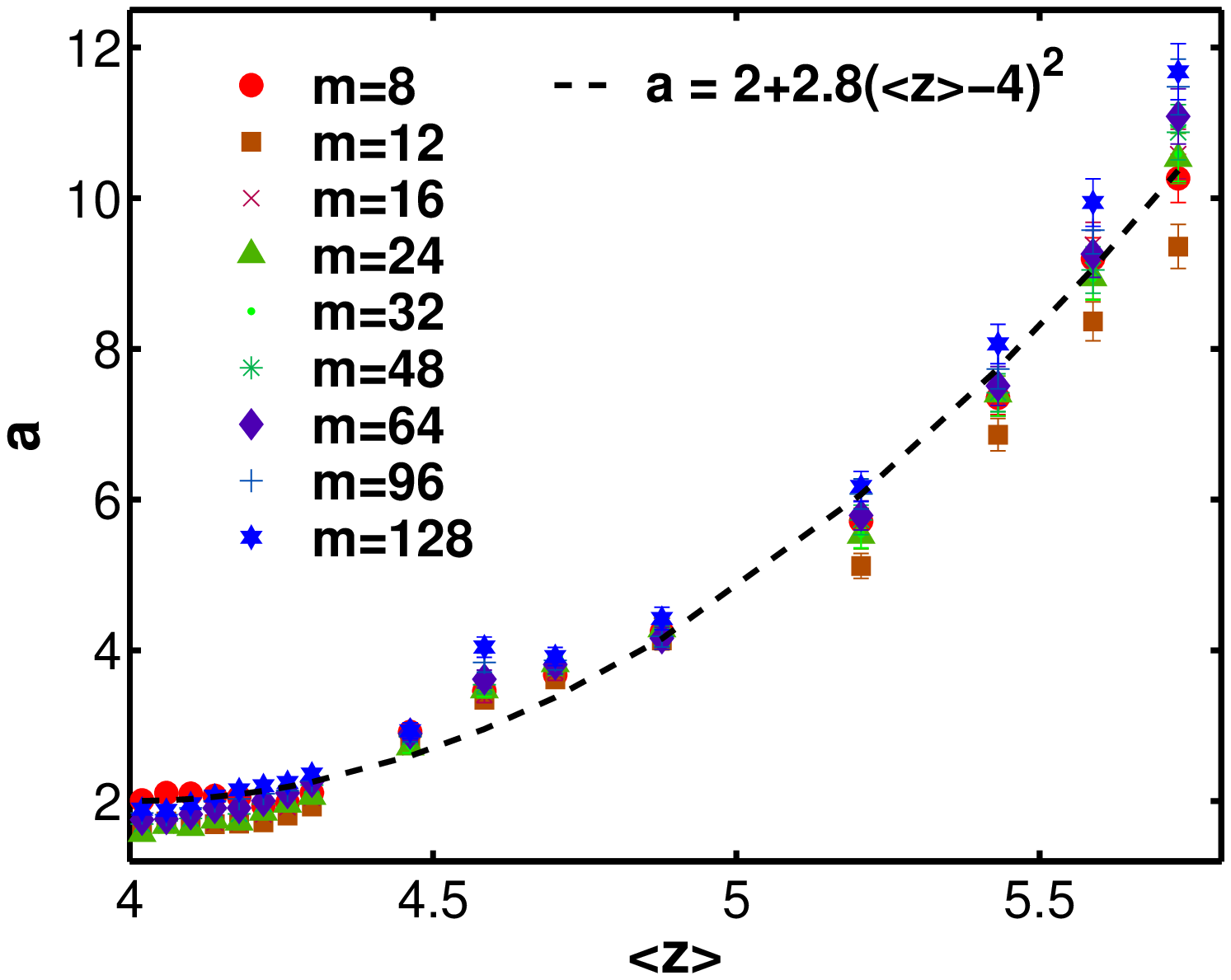}
\end{center}
\caption{Top: Distribution of $\Gamma$ for a subsystem size $m=24$ out of $N=4096$, and fit using equation \ref{omegasim} for the density of states. Bottom: Fit coefficient $a$ as a function of the contact number, with the fit which leads to equation \ref{omegasim}. \label{sim_figures}}
\end{figure}

For the systems with harmonic interactions,  we have also fitted the density of states $\Omega(\Gamma)$ to the form 
\begin{equation} \Omega(\gamma) \sim \Gamma^{m a} \quad \text{with} \quad a=2+c(z-z_{iso})^{2},\label{omegasim} \end{equation}
with $c = 2.8 \pm 0.5$ (see Figure \ref{sim_figures}).
Together with the thermodynamic relation $\alpha = \frac{\partial \ln(\Omega)}{\partial \Gamma}$ this form leads to an \emph{equation of state} over the full jammed range:
\begin{equation} \alpha = \frac{1}{\langle \Gamma \rangle}(2+c(\langle z \rangle-z_{iso})^{2}) \label{eqnstate}\end{equation} 
The numerical density of states shows deviations from the form consistent with equation \ref{eqnstate} for $m\leq 16$. It is likely that this is a more sensitive test of equation \ref{beta_fac}, and the correlation length $\xi \approx \sqrt{16}=4$ we can associate to this result is consistent with our results in the field theory section.

\subsection{Partition function at $z_{iso}$ \label{sec:Z_z_iso}}
The results deduced from simulations, equations \ref{omegasim} and \ref{eqnstate}, depend on the specific force law, much like the density of states of a solid depends on the detailed microscopic interactions. Close to the jamming transition $z=z_{iso}$, however, we can derive the form of the density of states $\Omega(\Gamma)$ and the equation of state $\alpha(\Gamma)$ by counting the number of degrees of freedom only. This indicates that the coarse-grained properties of packings of frictionless spheres are universal, i.e independent of force law and simulation protocol, as Point J is approached.

\paragraph*{Counting the states}
If we use the definition of the density of states, equation \ref{DOS_definition}, with equation \ref{can_distribution} in the canonical partition function equation \ref{can_scalar_partition}, we are able to write
\begin{equation} Z(\alpha) = \sum_{\nu} \beta_{\nu}^{dyn} \exp(-\alpha \Gamma_{\nu}), \label{ensemble} \end{equation}
where we sum over all packings consistent with force and torque balance, and with the force law respected (this last condition has to be modified for frictional packings). We only consider packings of frictionless spheres, so the torque balance constraint is automatically satisfied.

We can then formally separate the sum over configurations $\nu$ into a sum over all geometric configurations $\lbrace r_{ij} \rbrace$ and a sum over all force configurations $\lbrace F_{ij} \rbrace$, with $\delta$-functions to enforce the force-balance (f.-b.) and the force-law (f.-l.) constraints:
\begin{equation} Z(\alpha) = \sum_{\lbrace r_{ij} \rbrace} \sum_{\lbrace F_{ij} \rbrace} \beta_{\nu}^{dyn} e^{-\alpha \sum_{\langle ij \rangle} r_{ij} F_{ij}}\delta(\text{f.-b.})\delta(\text{f.-l.}) 
\end{equation}

For the frictionless packings, at the isostatic point, the number of degrees of freedom of the grains ($dN$) is equal to the number of forces that constrain them, $N \langle z \rangle /2$. Equating these gives the isostatic contact number, $z_{iso}=2 d$. It also shows that for a given geometry, there exists one and only one force configuration, and that the specific form of the force law becomes irrelevant. 

There is a one-to-one correspondence between a geometric configuration and a force configuration, so that we can eliminate the sum over the $\lbrace r_{ij} \rbrace$ and write:
\begin{equation} Z = \sum_{\lbrace F_{ij} \rbrace}  \beta_{\nu}^{dyn} \exp(-\alpha \sum_{\langle ij \rangle} r_{ij}(\lbrace F_{ij} \rbrace ) F_{ij}). \label{ZFr} \end{equation}
At the isostatic point, the mean force tends to zero, and the overlap (or deformation) of the grains becomes negligible. Therefore, the dependence of $\lbrace r_{ij} \rbrace$ on $\lbrace F_{ij} \rbrace$ can be neglected, $r_{ij} \rightarrow r_{0}$, assuming that the system is monodisperse for simplicity. 

The simplest measure $\beta_{\nu}^{dyn}$ with which the space of states is sampled we can choose is of course the flat measure $\beta_{\nu}^{dyn}=1$. Although not necessarily correct, this simple approximation allows us to treat the problem analytically and we are able to extract the correct density of states (see below).

If we choose a flat measure, no interaction terms between the different forces remain in equation \ref{ZFr}. Then the partition function can simply be written as a product:
\begin{equation} Z = \prod_{k=1}^{N z_{iso}/2} \int_{0}^{\infty} d F_{k} \exp(-\alpha r_{0} F_{k})= \left(\frac{1}{\alpha r_{0}} \right)^{N z_{iso}/2} \label{ZF} \end{equation}

\paragraph*{Connection to the simulation result}

The partition function can be rewritten as a function of the scalar force moment $\Gamma = p A \sum_{i=1}^{N z_{iso}/2} r_{i} F_{i}$. We insert this equation as an identity into equation \ref{ZF} and obtain after switching the order of integration:
\begin{equation*} Z(\alpha)=\int_{0}^{\infty} d\Gamma \frac{1}{r_{0}} e^{-\alpha \Gamma} \prod_{i=1}^{NZ_{iso}/2} \int dF_{i} \delta \left(\sum_{i=1}^{N z_{iso}/2} F_{i}-\frac{\Gamma}{r_{0}} \right). \end{equation*} 
We then perfom the force integral by enforcing the $\delta$-function and then integrating over the other remaining forces with the constraint $F_{N z_{iso}/2} \in (0,\Gamma/r_{0})$:
\begin{equation*} Z(\alpha)=\int_{0}^{\infty} d\Gamma \frac{e^{-\alpha \Gamma}}{r_{0}} \prod_{i=1}^{NZ_{iso}/2-1} \int_{0}^{\Gamma/r_{0}- \sum_{k=1}^{i-1} F_{k}} dF_{i}. \end{equation*} 
The remaining integrations describe the volume of a piece of the $Nz_{iso}/2-1$-dimensional hypercube, with volume $A_{d} = \frac{1}{d!} \left( \frac{\Gamma}{r_0} \right)^{d}$ \cite{Peskin}.

After dropping the prefactor the partition function is given by
\begin{equation} Z(\alpha) = \int_{0}^{\infty} \Gamma^{N z_{iso}/2-1}e^{-\alpha \Gamma}. \label{ZGamma}\end{equation}
This form is consistent with the simulation result for the density of states at the isostatic point ($z_{iso}=4$ in $2d$), $\Omega(\Gamma) = \Gamma^{2 m}$, in the limit $m>>1$. Either by using the thermodynamic relations $\alpha = \frac{\partial \ln(\Omega)}{\partial \Gamma}$ on equation \ref{ZGamma}, or $\langle \Gamma \rangle = -\frac{\partial \ln Z}{\partial \alpha}$ on equation \ref{ZF}, one can also obtain the universal equation of state
\begin{equation} \alpha = \frac{N z_{iso}}{2 \langle \Gamma \rangle}, \end{equation}
which matches equation \ref{eqnstate} for $\langle z \rangle \rightarrow z_{iso}$.
We have obtained the same equation of state for packings of disks with hertzian interactions close to $z_{iso}$ \cite{thesis}, showing that this result is independent of the interaction law.
The deviations from this form at small $m$ that we observe the simulations confirm that the assumption $\beta_{\nu}^{dyn}=1$ breaks down at very small scales. 

However, the agreement we obtain at larger $m$ shows that on a coarse-grained level, some of the properties of an isostatic packing can be understood through a simple model which assumes a flat measure in configuration space. Counting the number of degrees of freedom is then sufficient to explain the form of the equation of state and the density of states (basically, the ``thermodynamics'' of the system).

\paragraph*{Distribution of the forces}

Equation \ref{ZF} also predicts the single-force distribution in the canonical ensemble if we assume a flat measure $\beta_{\nu}^{dyn}=1$. From the form of the equation, we see that the probability to find a given force configuration is $P( \lbrace F_{1} ... F_{N z_{iso}/2} \rbrace) = \prod_{k=1}^{N z_{iso}/2} \exp(-\alpha r_{0} F_{k})$. Since this is a pure product distribution, we deduce the single-force distribution 

\begin{equation} P(F) \sim \exp(-\alpha r_{0} F) \sim \exp\left(-\frac{ziso}{4}\frac{F}{\langle \langle F \rangle \rangle}\right) \label{fdis_exp} \end{equation}

i.e. a pure exponential. We have used the equation of state as well as the definition of the \emph{ensemble average} of the forces $\langle \langle F \rangle \rangle =  \langle \Gamma \rangle/N r_{0}$ in the second equality.

We emphasize that the canonical stress ensemble does not imply an exponential form for $P(F)$ \emph{except} at the isostatic point, if we assume a flat measure. An exponential distribution emerges at the isostatic point because for a flat measure the forces are independent random variables at this point. Similarily, in an ideal gas, the Boltzmann distribution, $e^{-\beta E}$, becomes an exponential distribution of energies because the energy $E$ is a sum of single particle energies. For interacting systems, $e^{-\beta E}$ does not imply an exponential distribution of \emph{single particle} energies. 

Equation \ref{fdis_exp} is not a robust prediction since any deviation from the flat measure will have an especially strong effect on a single-particle quantity like the force distribution. With interactions, a calculation of $P(F)$ is challenging, as any one-body distribution is difficult to calculate for interacting systems\cite{Plischekebook}. The statistical mechanical framework that we lay out in this paper is much more amenable to calculating correlation functions and response functions, and this is the task we focus on in this paper.

A situation in granular materials that is very different from thermal systems is that numerically or experimentally, we do not have access to the canonical force distribution. Instead, by rescaling the force distributions by the \emph{spatial mean} $\langle F \rangle =\Gamma/ N r_{0}$ of the forces (instead of the unknown ensemble average $\langle \langle F \rangle \rangle$), we can measure the \emph{microcanonical} force distribution. O'Hern et al. established the algorithm used in our tests \cite{prl07} in \cite{epitome}, where Figure 16 (top) shows the microcanonical force distribution that can be obtained for the simulated system. It clearly decays faster than exponential. It is tempting to use the results derived above for the flat measure and to translate them to the microcanonical ensemble. However, in doing so, we would neglect all correlations between individual forces which are clearly important at the single-grain level.

\paragraph*{Other theoretical approaches}
Kruyt and Rothenburg \cite{Kruyt_Rothenburg} and Metzger et al. \cite{Metzger, Metzger2} use a maximum-entropy approach with a multi-component Lagrange multiplier very similar to $\hat{\alpha}$ to enforce that the total stress is conserved, and so work in the canonical $\alpha$ ensemble as well. The authors assume a product distribution for the forces and calculate the force distribution \emph{given} the distribution of contact angles and distances between grains. The result for the normal forces decays faster than exponential.

Another approach to the problem is the force network ensemble (FNE) introduced by Snoeijer et al. \cite{vanHecke_prl}, which uses the decoupling of forces and the positions of the grains for very stiff grains. For a hyperstatic packing with $\langle z \rangle > z_{iso}$, the FNE is then a microcanonical ensemble which assumes that for given a mean force $\langle F \rangle$ and a given geometry, all the configurations of positive compatible with force and torque balance are equally likely.

Tighe et al. \cite{Socolar} simulate the FNE on the strongly hyperstatic $\langle z\rangle=6>z_{iso}$ triangular lattice ad find a force distribution that decays faster than exponential for a system under isotropic compression, but an exponential decay for a sheared system. Recent work by Tighe et al. \cite{brian_leiden} introduces a second conserved quantity based on the height field to obtain a gaussian tail for an isotropic system.

Snoeijer et al. \cite{vanHecke} derive an analytical force distribution in the FNE for an isotropically stressed triangular lattice, as well as for a general geometry. They obtain a density of states which scales as $\langle F \rangle^{D}$, where $D\sim N(\langle z \rangle-z_{iso})$ is the number of excess force degrees of freedom in the system.

\paragraph*{Experimental results}
Measuring the force distribution inside a packing of grains is a challenge. Only two methods have so far been successful: 

Majmudar et al. \cite{Behringer_Nature, Trush} use quasi-twodimensional packings of photoelastic disks between cross polarizers and extract the stresses from optical measurements. For isotropic compression, they find that the distribution of the normal force components decays faster than exponentially, while the tangential force components follow an exponential distribution. If the system is subjected to pure shear, the distribution of the normal forces aquires an exponential tail while the tangential forces are not affected. The measurement is scaled by the spatial mean $\langle F_{n} \rangle$ of the normal forces, which relates this result to the microcanonical $\Gamma$-ensemble, as explained above in the context of the simulated data.

Brujic et al. \cite{Brujic_Makse_Colloid} as well as Zhou et al. \cite{Dinsmore} have measured the interparticle forces using confocal microscopy on index matched suspensions of droplets. Again, the results are given scaled by the mean force in the configuration. Brujic et al. have evidence for an exponential tail in the force distribution, while Zhou et al. focus on quantifying force chains.

Experiments on quasistatically sheared systems in a Couette geometery have also produced force distributions \cite{Jaeger_Nature05, Howell}, however the theoretical results above do not apply to dynamical systems. 

%***************************************************************************************************************************

\section{Building a field theory \label{sec:field_theory}}

By considering a field theory, we take a different route from the approach generally taken in the continuum mechanics community. The focus there is to find a \emph{constitutive relation} which links the stresses to the microscopic geometry of the packing. There is a considerable body of work on the subject in the mathematical and engineering literature (see e.g. \cite{Tordesillas} and references therein). The authors of \cite{BB} derive a constitutive relation for isostatic packings, though it can only be expressed at a microcopic level.

A field theory, however, uses a path integral over all the possible configurations for stable packings. It coarse-grains the microscopic details of each packing into a continuous field which is sufficient to describe the macroscipic properties of that packing. Then we only need to combine symmetry arguments with a perturbative expansion in the flucutations of the field around its mean to obtain the weight of a configuration in the path integral.

Here, we calculate correlations of the stress based on a minimal field theory that takes into account the essential features of a frictionless granular packing. The field theory is dominated by a laplacian leading term which is multiplied by a \emph{stiffness} which controls the behavior of the system as jamming is approached. We discuss the implications for the jamming transition. The field theory also predicts the correlation functions of the shear, which we test on simulated data.

This field theory is related to our earlier proposal for a field theory \cite{prl05} but, in this work, we have focussed {\it only} on stress correlations, and we have reassessed some of our assumptions based on information from simulations and experiments.  The predictions from the earlier field theory related pressure to deviation from isostaticity, and its predictions for the jamming transistion have been fit to experimental data by Behringer et al. \cite{Trush}.

\subsection{The Airy stress function}
To write down a field theory of granular packings, we need to identify a field which incorporates as many of the constraints placed on the system as possible, such that they do not have to be imposed separately. The intuitive choice of the pressure $p(x)=\Gamma(x)/A(x)$ as a field is misleading, since force and torque balance are not guaranteed for all possible configurations $p(x)$. Instead, we use the \emph{Airy stress function} $\Psi$, which incorporates force and torque balance constraints in two dimensions \cite{BB,Gurtin,thesis}, and which is related to the stress by
\begin{equation} \hat{\sigma}(x) = \vec{\nabla} \!\times \!\vec{h} = \vec{\nabla} \!\times \!\vec{\nabla} \!\times \!\Psi. \quad (\text{i.e.} \: \sigma_{ij} = \epsilon_{ik}\epsilon_{jl} \partial_{k} \partial_{l} \Psi), \label{Airy}\end{equation}
such that the pressure is given by $p = \text{Tr}(\hat{\sigma}) = \nabla^{2}\Psi$. 
The Airy stress function has been widely used in studying 2D elasticity, and especially the role of defects\cite{Chaikin_Lubensky}. In its traditional usage, $\Psi$ is obtained by minimizing the elastic energy. As will be seen from our analysis below, the field theory presented here has an entropic basis, and the role of $\Psi$ is very different.
After expressing the path integral in function of $\Psi$, the only remaining constraint is then that for purely repulsive granular packings, the local pressure has to be strictly positive.
For three dimensional systems, $\Psi$ has to be replaced by a $2$-tensor $\hat{\Psi}$ known as the Beltrami stress tensor \cite{Gurtin}. We do not consider this case here.

\subsection{Minimal field theory}
We will work in the microcanonical ensemble, where the total stress $\Gamma$ of the system and its contact number $z$ are fixed. The key quantity to predict is therefore the microcanonical partition function $Z(\Gamma,z)$ which is related to the canonical $Z(\alpha)$ by
\begin{equation} Z(\alpha) = \int d\Gamma dz \quad Z(\Gamma,z) \exp(-\alpha\Gamma). \end{equation}
In a first step, we limit our investigations to two-dimensional isotropic packings under pure compression, such that the total force moment tensor $\hat{\Sigma}$ can be written as $\hat{\Sigma} = \frac{\Gamma}{2} \hat{I} $.

Let $\psi$ be the deviation of the Airy stress function $\Psi$ from the one for a system with uniform pressure $p=\Gamma/A$. Then the local stress tensor can be entirely written as a function of the second derivatives of $\psi$
\begin{equation} \hat{\sigma} = \frac{\Gamma}{2 A} \hat{I} + \hat{\delta\sigma}=\frac{\Gamma}{2 A}\!\left(\!\! \begin{array}{cc} 1 & 0 \\ 0 & 1 \end{array} \!\!\right) + \left(\!\!\begin{array}{cc} \partial_{y}^{2} \psi & -\partial_{x}\partial_{y}\psi \\ -\partial_{x}\partial_{y}\psi & \partial_{x}^{2} \psi \end{array} \!\!\right)\!. \label{psi_fluct} \end{equation}
From equation \ref{Airy} we see that the Airy stress function admits a ``gauge invariance'' of the form
\begin{equation} \psi(x,y) \rightarrow \psi(x,y)+ax+by+c; \end{equation}
that is we have the freedom to choose two arbitrary constants while constructing the microscopic $\psi$ of a packing: the position of the origin for the fluctuations of $\vec{h}$ and $\psi$.
This means that all physical quantities need to be at least second derivatives of $\psi$.
The field theory has to honor this symmetry and therefore can only contain terms with at least second order derivatives of $\psi$.

Here, we consider only systems under an isotropic compression, and hence the system has to be (statistically) isotropic. All the terms in the action have to honor this symmetry as well. Then we write (the $\Gamma$ and $z$ dependence is through the coefficients $A$ and $B$)
\begin{equation} Z(\Gamma,z)=\!\!\int\!\! D\psi \exp \left[\!-\!\!\int\!\! dx dy \:A \text{Tr}(\delta\hat{\sigma})^{2}\! +\! B\text{Tr}(\delta\hat{\sigma}^{2}) \!+\! ...\right]\!. \label{field_formal}\end{equation}
In terms of the Airy stress function the two leading terms are given by $A (\nabla^{2}\psi)^{2} +B\left((\partial_{x}^{2} \psi)^{2}+(\partial_{y}\psi)^{2}+2(\partial_{x}\partial_{y}\psi)^{2} \right)^{2}$.

To extend this formalism to anisotropic systems, we need to decompose the stress tensor into a bulk term and the deviatoric stress $\delta\sigma_{ij}^{\text{dev}} = \delta\sigma_{ij}-\frac{1}{d}\delta_{ij} \delta\sigma_{kk}$. The coefficients multiplying the terms in the action are then analogous to the bulk and shear modulus of elasticity theory \cite{Field_Shear_draft}.

The similarity of the above to the free energy formalism one writes in elasticity theory \cite{Landau, Chaikin_Lubensky} is due to the same combination of a tensorial quantity and symmetry arguments. The two are in fact fundamentally different: first of all, elasticity theory is written as a function of the strain from a given reference geometry. The elastic free energy is then used to determine the properties of the system if the reference geometry is disturbed. For a granular system, there is no reference geometry and the strain is ill defined. Instead, equation \ref{field_formal} sums over all possible geometries and forces at a given $(\Gamma,z)$. Moreover, while in an elastic system the Lam\'{e} coefficients as well as the bulk and shear modulus are material constants, in this formalism they crucially depend on the imposed stress (see below). There is strong evidence that the bulk and shear modulus for granular and related systems depends on the imposed pressure and shear stress \cite{epitome}.

In Fourier space, all the lowest-order terms are proportional to $q^{4} |\psi_{q}|^{2}$, so that we can condense $A$ and $B$ into one coefficient $K$ and we write
\begin{align}
 Z(\Gamma,z) = & \!\int\!\! D\psi \exp \left[\!-\!\!\int\!\! \frac{d^{2}q}{2\pi^{2}} \left(K\!+\!A_{2}q^{2} \!+\!A_{4}q^{4}\!+\!...\!\right)|\psi_{q}|^{2} \right. \nonumber \\
& \left. +\lambda(q)|\psi_{q}|^{4} +... \right]. \label{field_full}
\end{align}

We have only investigated second-order correlation functions of the different components of the stress tensor in $q$-space. Since any eventual fourth order interaction terms $\lambda(q)|\psi_{q}|^{4}$ (or higher) will just renormalize the $A_{k}$ coefficients in the second-order correlation functions \cite{Peskin}, we are unable to probe them. Assuming possibly renormalized coefficients,
we obtain a correlation function for the fluctuations of the local pressure $\delta p =p-\Gamma/A = \nabla^{2} \psi$:
\begin{equation} \langle |\delta p_{q}|^{2}\rangle = q^{4}\langle |\psi_{q}|^{2}\rangle = \frac{1}{A_{0} +A_{2}q^{2}+A_{4}q^{4}+...} \label{Sqminimal}\end{equation}

\subsection{Simulation results}
We investigate the correlation functions $S(q)$ of the local pressure $p_{g} = \frac{1}{A_{g}}\sum_{i=1}^{z{g}} r_{gi} F_{gi}$ to find out if they are of the type expected from equation \ref{Sqminimal} and to determine the dependence of the coefficients on $\Gamma$ and $z$, as well as their interpretation. To obtain the data, we interpolate the discrete $p_{g}$ onto a grid of size $\sqrt{N}\times\sqrt{N}$ and then take the two-dimensional $FFT$ of the field. We then calculate the two-dimensional structure factor $|\delta p_{\vec{q}}|^{2}$ on the two-dimensional $q$-grid and finally take a radial average.

\begin{figure}[h]
\begin{center}
\includegraphics[width = 0.8\columnwidth]{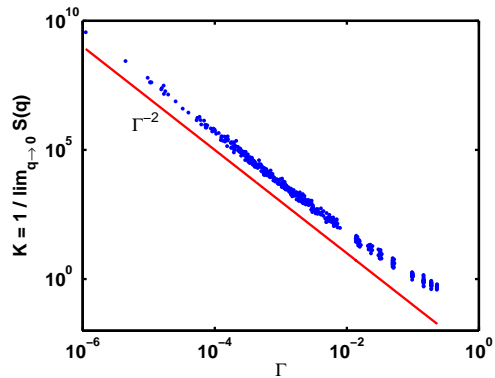}
\end{center}
\caption{Parameter $K$ for the $q$-independent term, determined from $K=1/\lim_{q\rightarrow0}S(q)$. The scaling $K\sim1/\Gamma^{2}$ is shown in red. \label{Kgam2}}
\end{figure}

\begin{figure}[h]
\begin{center}
\includegraphics[width = 0.7\columnwidth, trim = 0mm 0mm 15mm 0mm, clip]{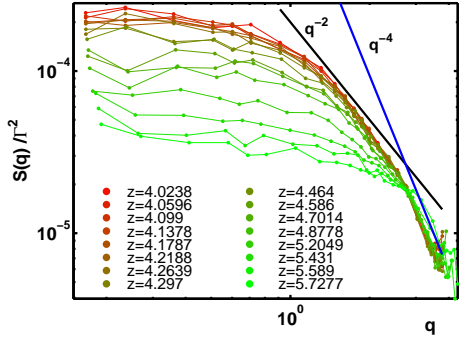}
\end{center}
\caption{Structure factor in log-log axes for systems with $N=1024$ grains, grouped by $z$. The $z$ increases from red to green, ranging from $z=4.02$ for the upper curve to $z=5.72$ for the lowest curve.}
\label{struplot}
\end{figure}

The first observation is that the structure factor has an overall scaling form
\begin{equation} S(q) = \Gamma^{2}s(q), \end{equation}
Figure \ref{Kgam2} shows the limit $K = \lim_{q \rightarrow 0} S(q)$ from the lowest $5$ radial $q$-points (see figure \ref{struplot}) as a function of $\Gamma$, with a $\Gamma^{-2}$-scaling over $4$ orders of magnitude. We only observe deviations in the limit of large $\Gamma$ (and $z$), far away from the jamming transition.
To test if $\langle |\delta p_{q}|^{2}\rangle$ has the form equation \ref{Sqminimal}, we rescale by $\Gamma^{2}$ and investigate the radial $s(q)$, which is then only parametrized by $z$. We group configurations with similar $z$, and average over the $s(q)$ to improve statistics (typically, about $20
$ configurations are averaged over). With $N=1024$, and a system size of $L\times L$ grain diameters ($L=32-37$, depending on packing fraction), we can investigate the range of $q$ from $ \frac{2\pi}{L}$ to $\frac{\sqrt{2}N_{\text{grid}}}{2} \frac{2\pi}{L}$, where $N_{\text{grid}}=32$ is the size of the grid we use for $\delta p$.

Figure \ref{struplot} shows the structure factor obtained for $N=1024$ for all $z$. The low-$q$ values of $s(q)$ decrease with increasing $z$, while the tail of the function does not change. The $q^{-2}$ and $q^{-4}$-lines provided as a guide to the eye help to show the smooth transition of $s(q)$ from a constant at low $q$ through a $q^{-2}$ decay at intermediate $q$ to $q^{-4}$ at high $q$. We were able to fit all the curves to equation \ref{Sqminimal} with $3$ fitting parameters, and the results led us to write the following form:
\begin{equation} s(q) = \frac{1}{k_{0}(a(z)+\xi_{2}^{2}q^{2}+\xi_{4}^{4}q^{4})} \label{sqform}\end{equation}

\begin{figure}[h]
\begin{center}
\includegraphics[width = 0.7\columnwidth, trim = 0mm 0mm 15mm 0mm, clip]{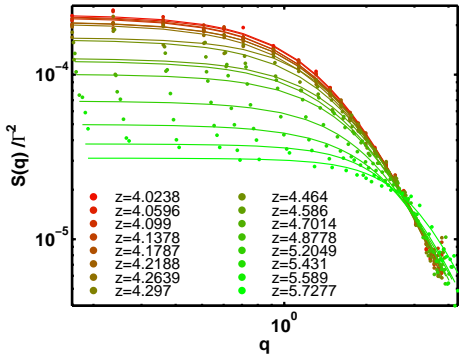}
\end{center}
\caption{Fit of the structure factor to the form equation \ref{sqform} for systems with $N=1024$ grains, grouped by $z$. \label{strufit}}
\end{figure}

\begin{figure}[h]
\begin{center}
\includegraphics[width = 0.7\columnwidth, trim = 0mm 0mm 15mm 0mm, clip]{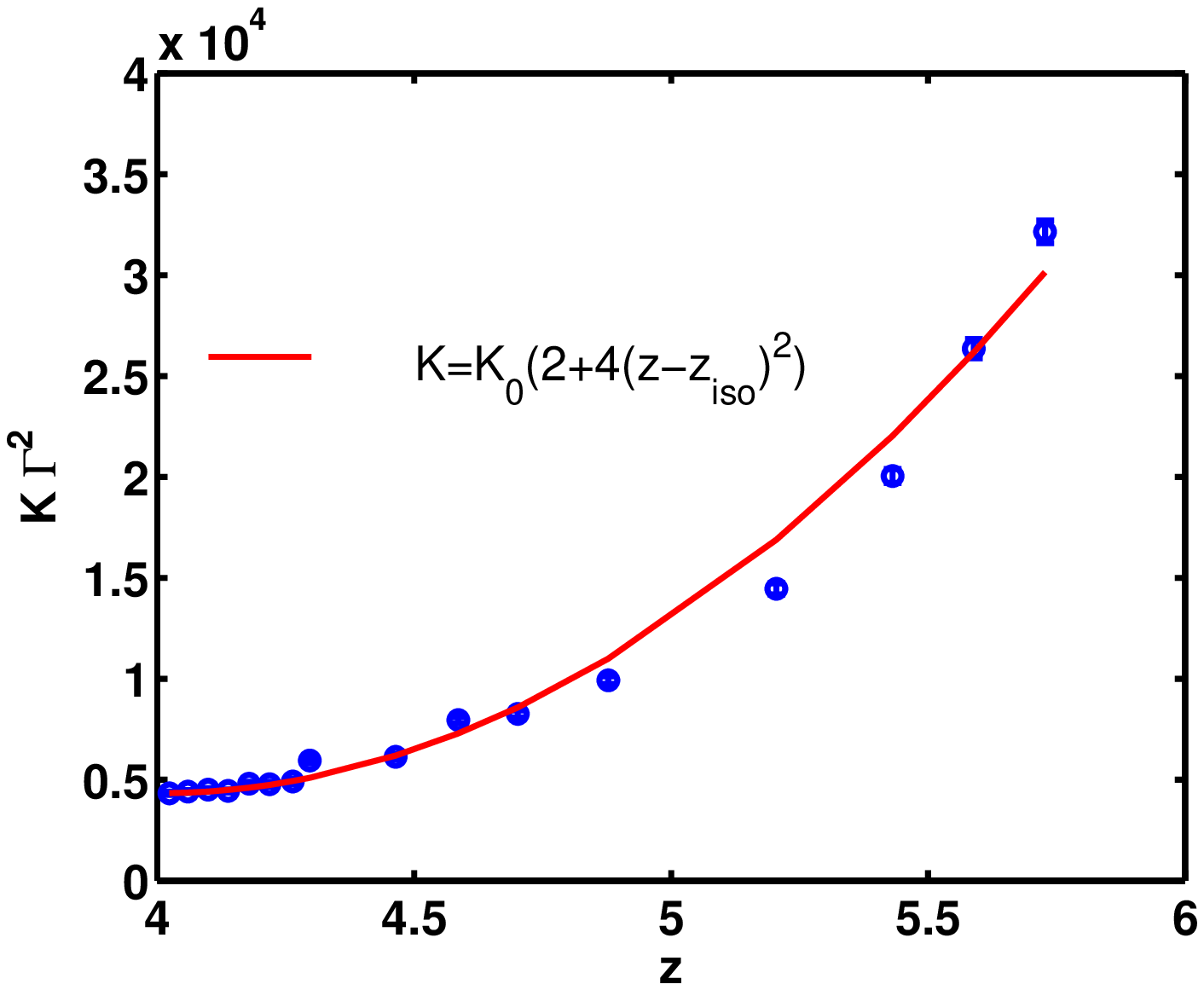}
\includegraphics[width = 0.7\columnwidth, trim = 0mm 0mm 15mm 0mm, clip]{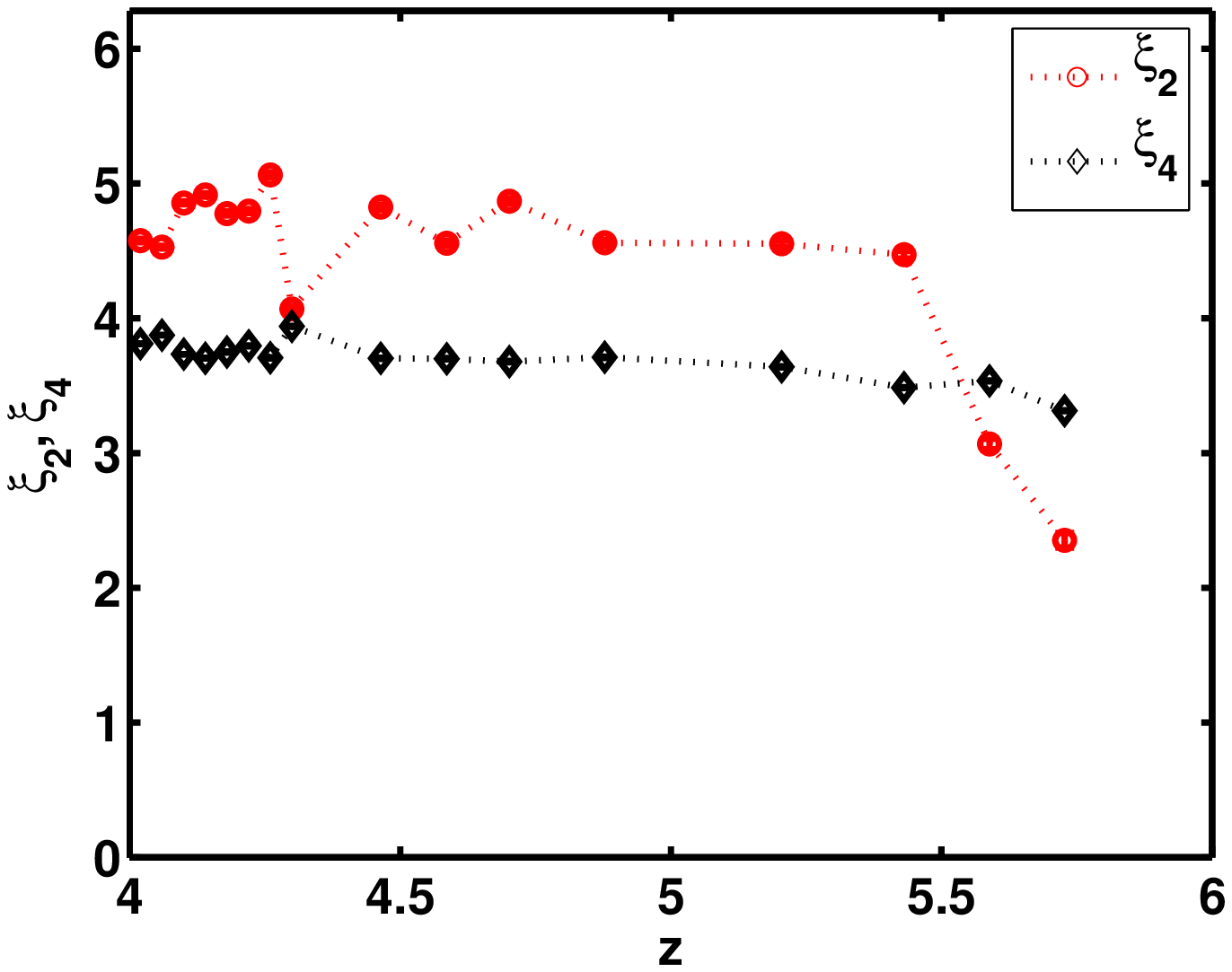}
\end{center}
\caption{a) Fitting parameter $K$ of the $q$-independent term, and fit to form equation \ref{sqform}. b) Length scales $\xi_{2}$ and $\xi_{4}$ associated to the $q^{2}$ and $q^{4}$-terms, in units of $a$.}
\label{kzfit_xi24}
\end{figure}

Figure \ref{strufit} shows the fitted curves, while Figure \ref{kzfit_xi24} a and b show the parameters $k=k_{0}a(z)$, $\xi_{2}$ and $\xi_{4}$.

The following picture emerges: The two length scales $\xi_{2}$ and $\xi_{4}$ are very nearly independent of the contact number $z$, and much smaller than the system size, with $\xi_{2} = 4.5\pm 0.5$ and $\xi_{4} = 3.75\pm 0.25$. This suggests that they reflect the purely microscopic properties of the system, like the distribution of grain sizes, which do not influence the behaviour of the system at larger scales. The $q$-independent term in the denominator, however, depends on $z$ and is responsible for the supression of the low-$q$ fluctuations with increasing contact number. The $z$-dependence is quadratic, 
\begin{equation} k(z) = k_{0}(2+\tilde{c}(z-z_{iso})^{2}) \quad \text{with} \quad \tilde{c}=4 \pm 0.5, \end{equation}
with a coefficient of the same order of magnitude as the $c$ of the equation of state \ref{eqnstate}.

The full second-order correlation function is then given by:
\begin{equation} S(q) = \frac{\Gamma^{2}}{K_{0}} \frac{1}{(2+\tilde{c}(z-z_{iso})^{2}) + \xi_{2}^{2} q^{2}+\xi_{4}^{4} q^{4}} \label{sq_sim}\end{equation}

\subsection{Origin of the stiffness}
The previous section has shown that the behavior of the system at large length scales is dominated by the first term in the field theoretical model, $|\nabla^{2}\psi|^{2}$. The length scales present in the system are just a few grain diameters, and remain small when the jamming transition is approached. There is no evidence of a growing static length scale in the system when the jamming transition is approached.
This is consistent with our observations in the ensemble section, where we found that for lengthscales larger than approximately $4$ grain diameters, the density of states factorizes and the approximation of a flat measure in configuration space becomes valid.
It is possible that there is a length scale unrelated to the pressure in the system, or a length scale that the finite size simulations are not sensitive to. Interestingly, the $\psi$ field is critical and has power-law correlations, independent of the distance from jamming (see also next section).

In the FNE, where by definition there are no correlations between the particles, it has been shown that the pressure correlations are flat, and that the correlations of the Airy stress function decay as $\sim 1/q^{4}$, consistent with a field theory with only the stiffness $K$ and no higher-order terms \cite{Brian_positivity}.

Our analysis raises many questions about the behavior of static granular packings near jamming (or more appropriately, unjamming).  It is surprising that the simulations show no crossover to usual elasticity theory, and that the $\Psi$ field remains critical even away from jamming.  Usual elasticity theory would lead to a unique solution for $\Psi$ at a given $\Gamma$ and $z$.

Can we understand why the form of the field theory is so very different from the common universality classes encountered in statistical mechanics? For example, in the $\phi^{4}$-model the phase transition occurs when the mass term vanishes compared to the gradient term and higher-order terms \cite{Chaikin_Lubensky}. In our case, the mass term is absent because of the gauge invariance, and the $\psi$-field is always critical.

We need to understand the coefficient of $(\nabla^{2}\psi)^{2}$,  
\begin{equation} K(z) = \frac{K_{0}}{\Gamma^{2}}(2+\tilde{c}(z-z_{iso})^{2})  \label{Kz} \end{equation}
and how it is related to the jamming transition at $(\Gamma=0,z=0)$.

We can predict the scaling of the \emph{stiffness} $K(\Gamma,z)$ with $\Gamma$ if we take into account the constraint that the local pressure $p(r)$ has to be positive for all $r$. The argument below was originally proposed by B. Tighe in the context of the force network ensemble \cite{Brian_positivity}.

Let $p(r) = \Gamma/A + \nabla^{2} \delta \psi >0$. After transforming into Fourier space, this condition becomes:
\begin{equation} \int \frac{d^{2} q}{(2 \pi)^{2}} \quad q^{2} \psi_{q} e^{i \vec{q}.\vec{r}} \quad \leq \Gamma/A \end{equation}
The LHS ($=-\nabla^{2}\psi(r)$) is the negative of the local deviation from the mean pressure and can be both positive or negative. If the LHS is negative, the local pressure fluctuation is positive, and the constraint is automatically fulfilled. We therefore consider the case where the LHS is positive. We can square both sides while keeping the inequality, and transform one of the integrals by noting that $\psi=\psi^{*}$
\begin{equation*} \int \!\frac{d^{2} q}{(2 \pi)^{2}} \: q^{2} \psi_{q} e^{i q r} \int \!\frac{d^{2} q'}{(2 \pi)^{2}} \: q'^{2} \psi_{q'}  e^{-i q' r} \:\leq \left(\frac{\Gamma}{A}\right)^{2}\!. \end{equation*}
If we integrate over all $r$ on both sides, the $RHS$ aquires a volume term  $A$, while in the $LHS$, we can change the order of integrations and get a $(2\pi)^{2}\delta(\vec{q}-\vec{q'})$ from the exponentials. The condition becomes then:
\begin{equation*} \int \!\frac{d^{2} q}{(2 \pi)^{2}} \: q^{4} |\psi_{q}|^{2} \: \leq \Gamma^{2}/A\!. \end{equation*}
The integrand is always positive, so we can write:
\begin{equation} q^{4} |\psi_{q}|^{2} \leq \left(\frac{\Gamma}{A}\right)^{2} \end{equation}
The $LHS$ is nothing but $S(q) = 1/K(\Gamma,z)$ and so, finally, the positivity constraint leads to 
\begin{equation} K(\Gamma,z) \geq \frac{A^{2}}{\Gamma^{2}} \end{equation}
The field theory we have constructed is for the marginal case where the stiffness satisfies the equality, and the stiffness is therefore the smallest allowed by the constraint of positivity.
The $z$ dependence of $K$ is nontrivial and not predicted this argument. Its form, however, is not totally unexpected: At larger contact numbers, the stiffness, which is related to the number of configurations, should increase since there are more configurations available to the packing.

\subsection{Tests of the field theory}

\paragraph*{Implications for Jamming} 
The form of the field theory we have obtained from simulations and from theoretical arguments presents a picture of the jamming transition where the transition is the result of the number of possible states for the system tending to zero. As $\Gamma$ tends to zero, the stiffness diverges, and hence the number of states around the smooth-pressure ground state that the system can access under perturbations is drastically reduced. This is ultimately a consequence of the positivity constraint: the position of the hyperplane on which the $\Gamma$-constraint is satisfied shifts to the ``lower left corner'' of the space of allowed forces, so that the area of the hyperplane shrinks drastically.

\begin{figure}
\begin{center}
\includegraphics[width = 0.8\columnwidth, trim = 15mm 0mm 15mm 0mm, clip]{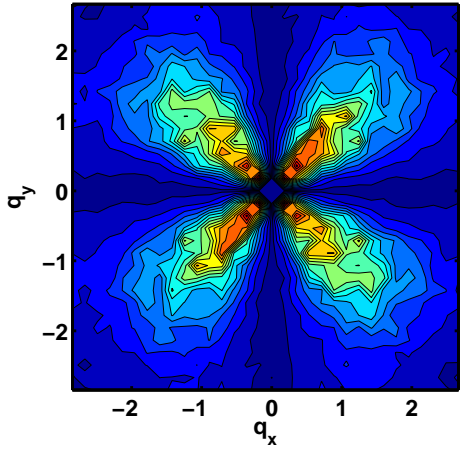}
\end{center}
\caption{Structure factor $\langle |(\hat{\sigma}_{xy})_{q}|^{2} \rangle$ obtained from the mean of $20$ simulated packings at $\langle z \rangle = 5.73$. \label{shear_struc_sim}}
\end{figure}

\begin{figure}
\begin{center}
\includegraphics[width = 0.8\columnwidth, trim = 15mm 0mm 15mm 0mm, clip]{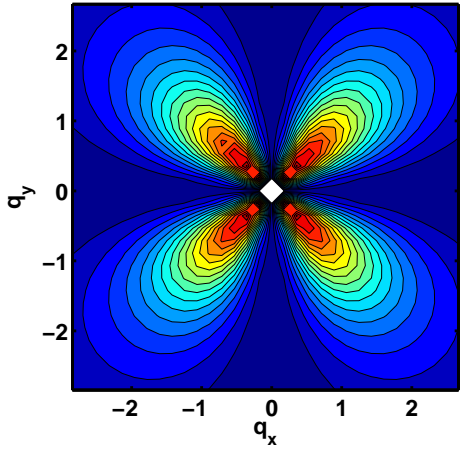}
\end{center}
\caption{Theoretical prediction of equation \ref{shear_strufac} for $\langle |(\hat{\sigma}_{xy})_{q}|^{2} \rangle$, at the same $\langle z \rangle$. \label{shear_struc_theo}}
\end{figure}

\paragraph*{Fluctuations of the shear stress}
Even for isotropic systems, where the global shear stress $\hat{\Sigma}_{xy}/A$ is zero, the local values of $\hat{\sigma}_{xy}$ are nonzero and have well-defined correlations. Since the shear can be expressed as a function of the Airy stress function as $\hat{\sigma}_{xy} = -\partial_{x}\partial_{y}\Psi$, we can predict the shear structure factor in Fourier space:
\begin{equation} \langle |(\sigma_{xy})_{q}|^{2}\rangle = \frac{\Gamma^{2}}{k_{0} q^{4}} \frac{q_{x}^{2}q_{y}^{2}}{(2\!+\!c(z\!-\!z_{iso})^{2}) + \xi_{2}^{2} q^{2}+\xi_{4}^{4} q^{4}}.  \label{shear_strufac}
\end{equation}
Figure \ref{shear_struc_sim} shows the structure factor of the shear obtained from the mean of $20$ simulated packings, while the theoretical prediction from equation \ref{shear_strufac} is shown in Figure \ref{shear_struc_theo}. 

We find good agreement between the simulation result and the prediction, especially for the angular structure of the correlation function.

\paragraph*{Real-space correlation functions}
From the form of the Fourier-space correlation functions equations \ref{sq_sim} and \ref{shear_strufac}, we can predict the real-space correlation functions e.g.,
\begin{equation} \langle \delta p(\vec{r}) \delta p(0) \rangle = \int \frac{d^{2} q}{(2\pi)^{2}} \exp(i \vec{q}.\vec{r}) \langle |\delta p_{q}|^{2} \rangle \label{prcorr_form} \end{equation}
and determine if the field admits long-range correlations. We condense the short correlation lengths $\xi_{2}$ and $\xi_{4}$ into a single correlation length $\xi\approx 4 $, which multiplies the $q^{2}$ term. This is a good approximation since the integrals are dominated by the small-$q$-limit, where the $q^{2}$-term dominates the $q^{4}$-term in the denominator.

\begin{figure}[h]
\begin{center}
\includegraphics[width = 0.75\columnwidth]{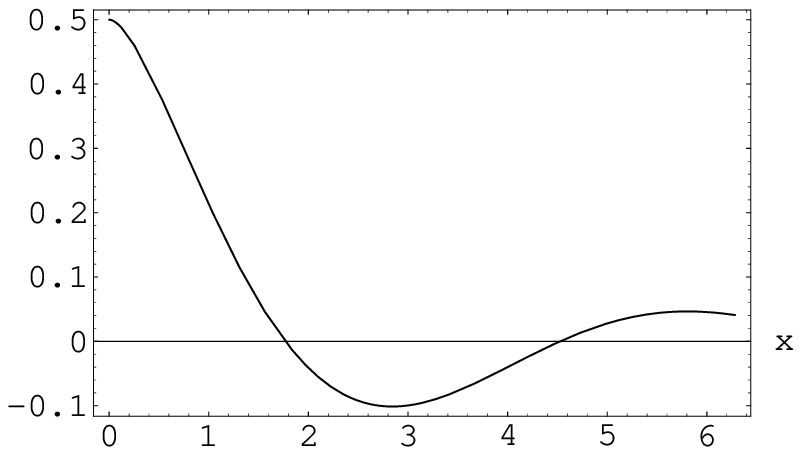}
\includegraphics[width = 0.7\columnwidth, trim = 10mm 0mm 10mm 5mm, clip]{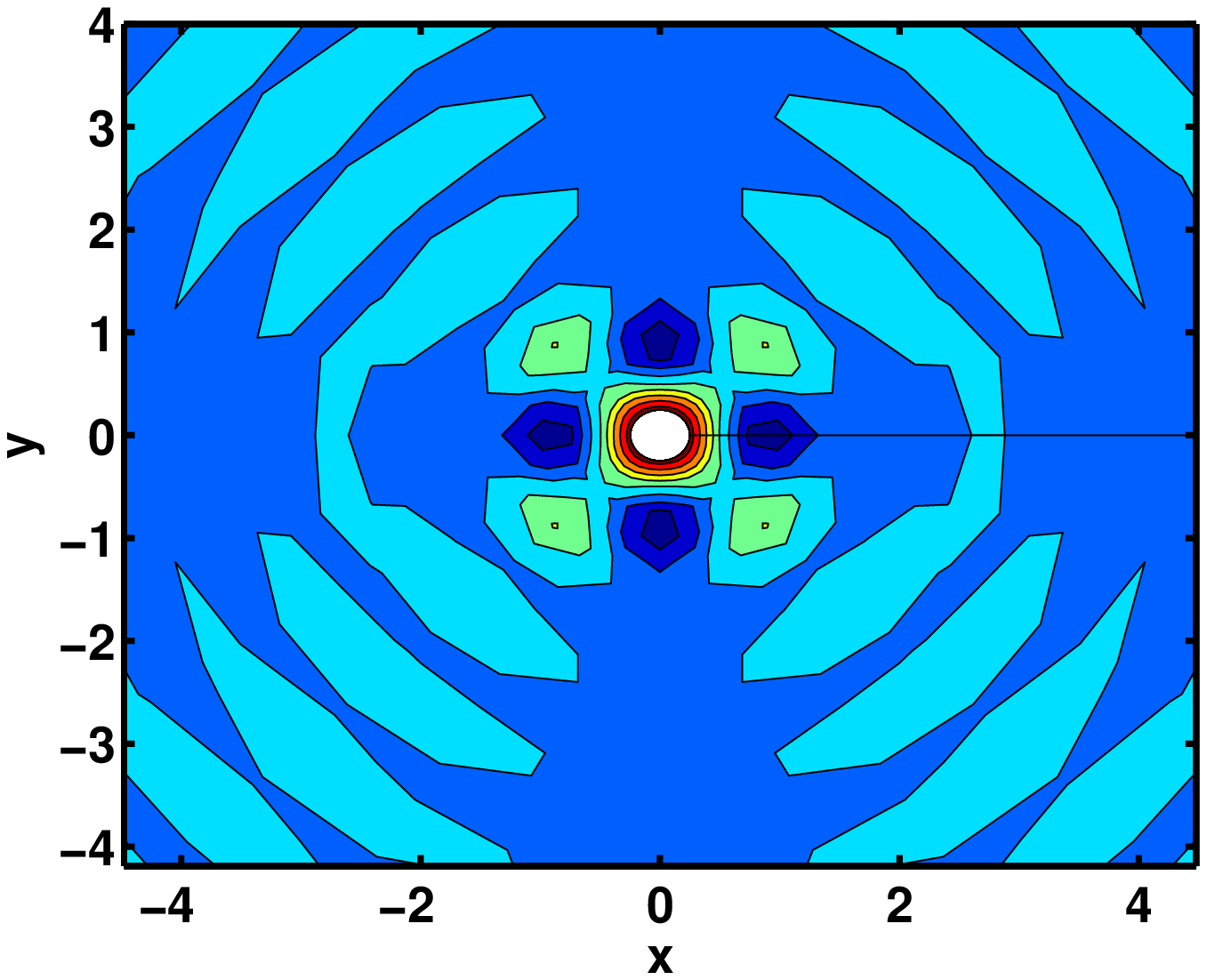}
\end{center}
\caption{Top: Real-space correlations of the Airy stress function, expressed as a function of the scaling variable $x=\frac{2\pi r}{L}$. Bottom: Real-space correlations of the shear stress, obtained by numerical integration of equation \ref{shear_strufac}.}
\label{shear_real_A}
\end{figure}

\paragraph*{Airy stress function}
The second-order correlation function of the fluctuations of the Airy stress can be calculated for $\xi\rightarrow 0$, if the integral is cut off at system size (the cutoff is necessary even for $\xi>0$). We obtain a scaling form in $x=\frac{2\pi r}{L}$,
\begin{align} &\langle \psi(\vec{r})\psi(0)\rangle = \frac{\Gamma^{2}}{K_{0}(2+\tilde{c}(z-z_{iso})^{2})} \frac{L^{2}}{128 (2\pi)^{3}} x^{2} \nonumber \\
 &\left[ - x^{2} _{2}F_{3}(1,1;2,3,3;-x^2)+32 \left(\ln(x/2)\!+\!2 x^{2}\!+\!\gamma\!-\!1\right)\right], \end{align}
where $_{2}F_{3}(1,1;2,3,3;-x^2)$ is a hypergeometric function and $\gamma$ is the Euler-Mascheroni constant. Figure \ref{shear_real_A} shows the correlation function for $r\in(0,L)$, and it is apparent that the correlations of the Airy stress function are long ranged and scale with the system size - qualitatively consistent with the real-space correlations seen in \cite{thesis}.

\paragraph*{Pressure fluctuations}
The inverse Fourier transform equation \ref{prcorr_form} can be performed analytically. After the angular integration we obtain
\begin{equation} \langle \delta p(r) \delta p(0) \rangle = \frac{\Gamma^{2}}{ K_{0}}  \int_{0}^{\infty} \frac{d q}{2 \pi} \frac{J_{0}(q r)}{2+c(z-z_{iso})^{2}+q^{2}\xi^{2}} \end{equation}
The result is another Bessel function, $K_{0}(x)$, the $0^{th}$ modified Bessel function of the second kind:
\begin{equation} \langle \delta p(r) \delta p(0) \rangle = \frac{\Gamma^{2}}{K_{0} 2\pi} K_{0}\left(\frac{(2\!+\!c(z\!-\!z_{iso})^{2})^{1/2} r}{\xi} \right)\!. \end{equation}
Asymptotically, for large arguments, $K_{0}(x) \sim e^{-x}/x^{1/2}$ so that the pressure fluctuations have short-range correlations which fall off beyond a scale set by the correlation length $\xi$. 
% Please, please, who did this first explicitely?
% This is consistent with \cite{??}, which do not observe long-range correlations of the pressure.

\paragraph*{Shear fluctuations}
For the shear fluctuations, the integral has a nontrivial quadrupolar angular dependence and can only be done analytically for certain angular directions ($\theta = 0, \pi/2,\pi$ and $3\pi/2$) using polar coordinates. The result for these directions is a combination of Bessel functions and a constant piece, multiplied by a power law, 
\begin{align} 
&\langle \delta \hat{\sigma}_{xy}(r) \hat{\sigma}_{xy}(0) \rangle = \frac{\Gamma^{2}}{ K_{0}} \frac{2\pi}{r^{2}}\left[ \frac{1}{f(z)}\left \lbrace\!-\!\frac{1}{2}\!+\!K_{2}\left(\frac{f(z)^{1/2} r}{\xi} \right)\right \rbrace \right. \nonumber \\
&\left.-\frac{r}{f(z)^{1/2} \xi} K_{3}\left(\frac{f(z)^{1/2} r}{\xi} \right) +6\left(\frac{\xi}{f(z) r}\right)^{2}\right]\!, \label{shear_realspace}
\end{align}
where $K_{3}$ and $K_{3}$ are the second and third modified Bessel function of the second kind, and $f(z)=2+c(z-z_{iso})^{2}$.
The Bessel functions decay exponentially for $r>\xi$, however the first term shows that the shear admits long range, power-law correlations $\sim 1/r^{2}$, regardless of the distance from the jamming transition. Equation \ref{shear_realspace} has negative correlations, and in fact a numerical integration of equation \ref{shear_strufac} shows that the real-space correlation function has a similar quadrupolar angular dependence as the Fourier-space form, with negative corrlations along the axes, and positive correlations at $45^{o}$ (see Figure \ref{shear_real_A}; the oscillations in the correlation functions are the result of a sharp upper cutoff of the $q$-integral at grain size $q_{max} = \frac{2\pi}{a}$).

%***********************************************************************************************************************
\section{Mean field theoretical model}
The results from the simulation, as well as the universal properties we have found at the isostatic point can be combined into a mean-field theory of jammed packings of frictionless spheres. We investigate the properties of the minimum of the effective free energy $F$, and show that Point J has some properties of a critical point within this framework.

\subsection{An effective free energy}

\paragraph*{At the isostatic point}
The results for the density of states from the simulation and the partition function at the isostatic point we derived in section 2 agree with each other (equations \ref{omegasim} and \ref{ZGamma}).
We can define an intensive mean-field variable $\gamma = \Gamma/m$, such that $P(\gamma) = \gamma^{2m} e^{-\alpha m \gamma} = \left( \gamma^2 e^{-\alpha \gamma} \right)^{m}$. Then we write a free energy as a function of this variable:
\begin{align}
Z(\alpha) = & \int_{0}^{\infty} d \gamma e^{-m F(\gamma)} \quad \text{with}  \nonumber \\
& F_{z_{\text{iso}}}(\gamma) = \alpha\gamma-2 \ln(\gamma). 
\end{align} 

\paragraph*{Scaling of the mean contact number}
The variable which parametrizes the departure from the isostatic point is the mean contact number $\langle z \rangle \equiv z$, in mean-field notation. From the simulations, we were able to extract several scaling laws linking the contact number and $\gamma$, by exploring the phase space of compressed jammed configurations available to the conjuguate gradient minimization protocol. 

\begin{figure}[h]
\begin{center}
\includegraphics[width = 0.49\columnwidth, trim = 5mm 0mm 15mm 0mm, clip]{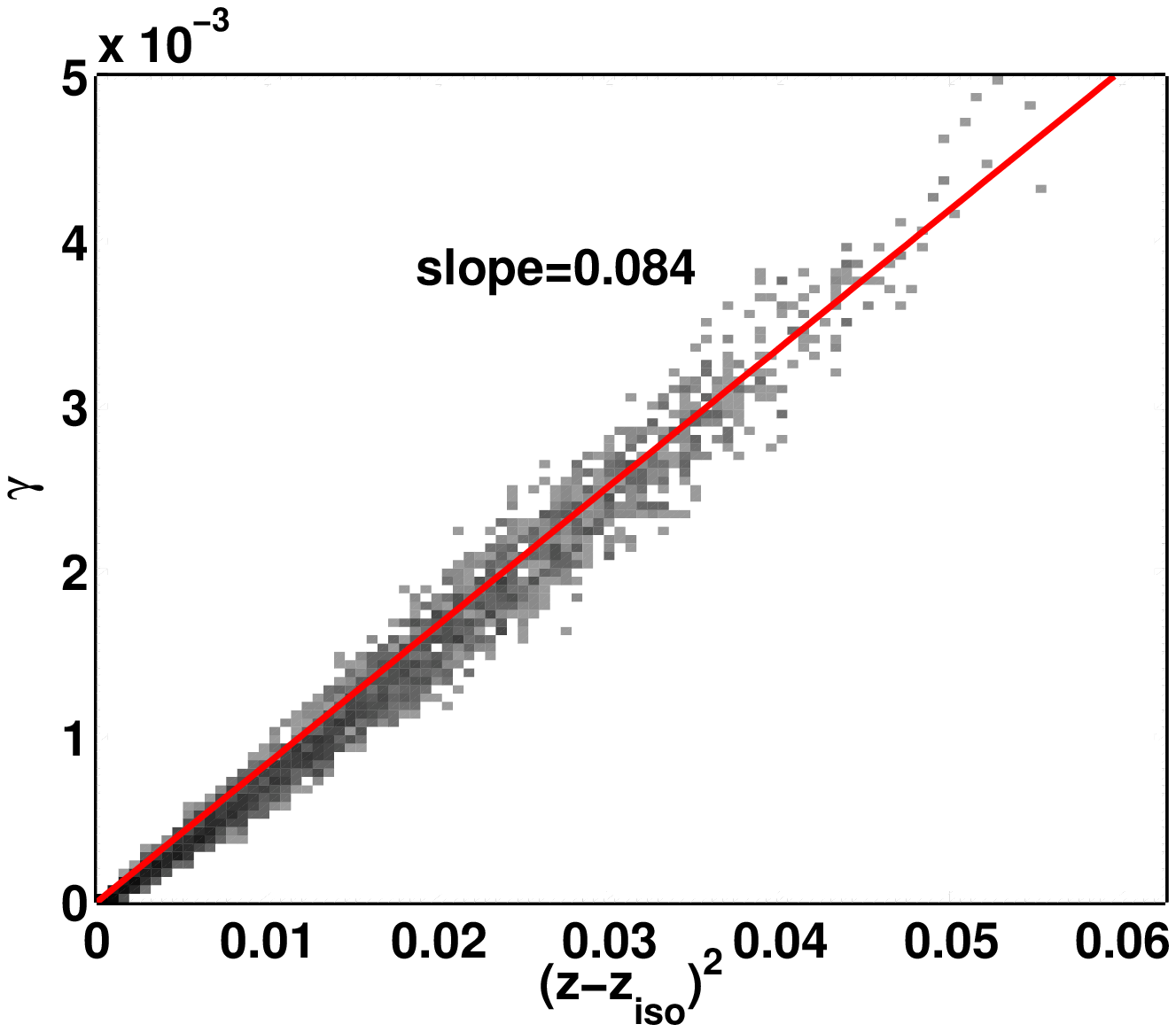}
\includegraphics[width = 0.49\columnwidth, trim = 5mm 0mm 15mm 0mm, clip]{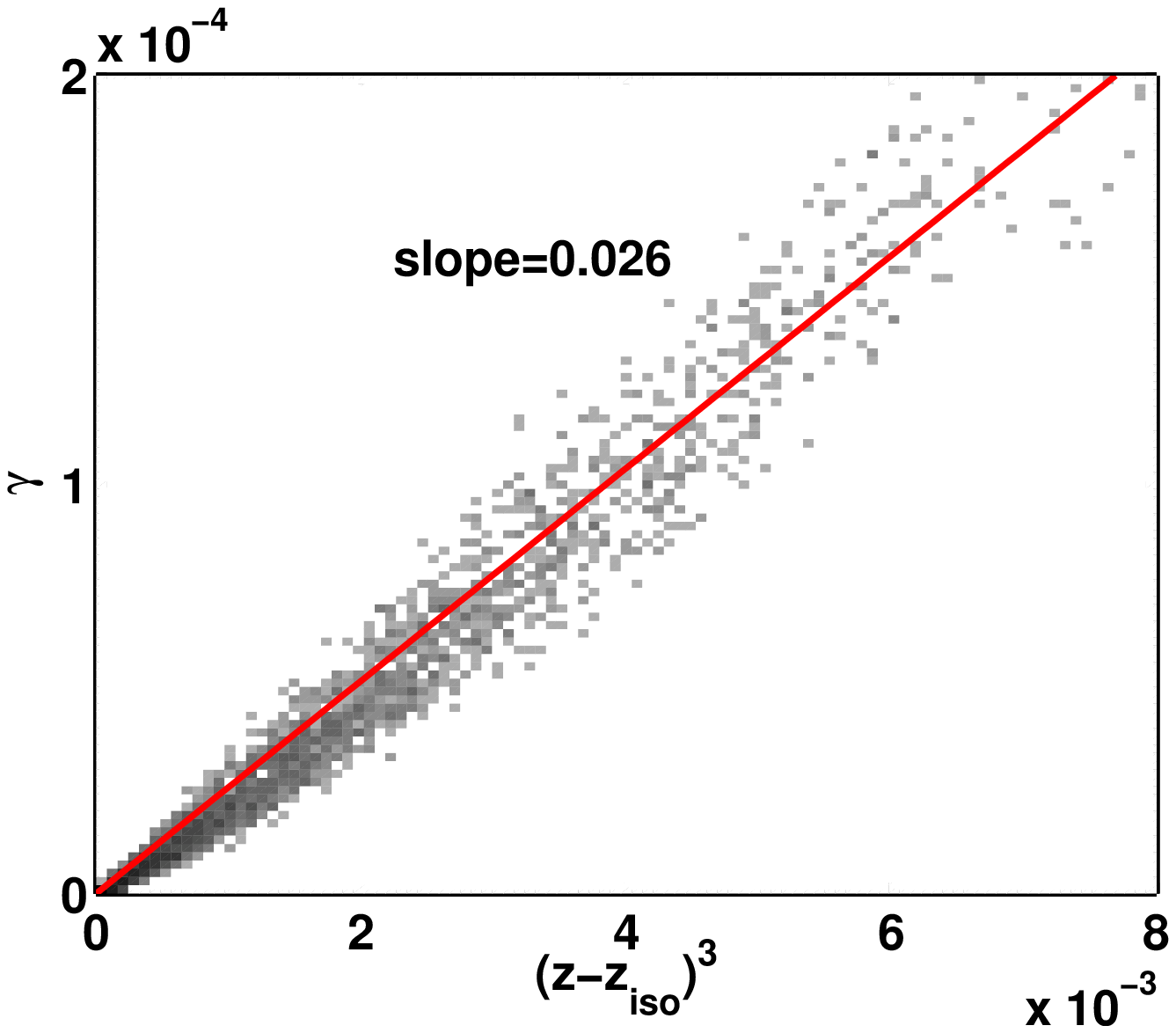}
\end{center}
\caption{Coefficient for the power law scaling between $\langle \gamma \rangle$ and $\langle z \rangle-z_{iso}$ for harmonic interactions (left) and hertzian interactions (right). \label{z2gamma}}
\end{figure}

We observe the following relation between the ensemble means of $\langle \gamma \rangle$ and $\langle z \rangle$ (see Figure \ref{z2gamma}),
\begin{equation} \langle \gamma\rangle = B_{s} (\langle z\rangle-z_{iso})^{2},\label{gamz2}  \quad \text{with} \quad B_{s}=0.084\end{equation}
\begin{equation} \langle \gamma \rangle = B_{h} (\langle z \rangle-z_{iso})^{2},\label{gamz3}  \quad \text{with} \quad B_{h}=0.026 \end{equation}
for harmonic and hertzian interactions, respectively. This scaling was first observed in the Chicago simulations \cite{epitome}, from which our simulation protocol derives.

This scaling, and by extension the empirical mean field free energy we present below, do depend significantly on the precise form of the intergranular potential (quadratic or hertzian in this case) and the nature of the simulation protocol. We explore the influence of the simulation protocol in the context of a stability argument. 

\paragraph*{Form of the free energy}
For $z>z_{iso}$, the number of additional force variables in a system of size $N$ is $N(z-z_{iso})/2$ and each geometric configuration can bear several force-balanced states. The form of the partition function at the isostatic point equation \ref{ZF} needs to be updated to take account of the additional force variables, and the geometrical information cannot be dropped:
\begin{equation} Z = \prod_{k=1}^{N z/2} \int_{0}^{\infty} d F_{k} \exp(-\alpha \gamma) \sum_{\lbrace r_{ij}\rbrace} \delta\left( \text{geom.-constr.} \right) \label{Zform} \end{equation}
The constraints imposed by the geometry on each arrangement of the forces are similar to the case studied by Snoeijer et al. \cite{vanHecke}. Their work focuses on integrating over all the possible force distributions allowed by force balance and purely repulsive forces on a given triangular lattice with fixed external forces. The case studied here is essentially an inversion of the problem: for a given set of forces, is there a geometry of ($N$, frictionless, spherical) particles to accommodate them? Their direct approach, even if conceptually feasible for a single random geometrical configuration, fails for us because we have no Ansatz to tackle the problem of counting all the possible geometrical configurations at large compressions. 

We assume that the formal equation \ref{Zform} is consistent with writing $Z$ as a function of $\gamma$ and $t=z-z_{iso}$:
\begin{equation}
Z = \int_{0}^{\infty} d\gamma \int_{0}^{\infty} d t \exp \left(m F_{z_{\text{iso}}}(\gamma)\right) \exp\left(m W(t,\gamma)\right),
\end{equation}
such that the effective free energy is $F(\gamma,t,\alpha) = \alpha\gamma - 2 \ln \gamma +W(t,\gamma)$. We then determine the simplest $W$ consistent with the simulation results discussed above. 

For a system with harmonic interactions, equation \ref{omegasim} directly leads to $W(t)=-c (t)^{2}\ln(\gamma)+g(t)$, and this also gives the correct equation of state \ref{eqnstate} from setting the first derivative $\frac{\partial F}{\partial \gamma}$ to zero. To incorporate the relation \ref{gamz2}, we set the $z$-derivative of $F$ to zero and substitute equation \ref{gamz2} for $\gamma$, so that we find
\begin{equation} g(t)= c t^{2} \left[\ln(B_{s} t^{2})-1\right]. \end{equation}
Then, the effective free energy is given by
\begin{equation} F(\gamma,z)=\alpha \gamma - 2\ln \gamma - c t^{2} \left[\ln\left(\frac{\gamma}{B_{s}t^{2}}\right)+1\right]\label{free_empirical_spring} \end{equation}

For systems with hertzian interactions, we only know the relation \ref{gamz3} between $\gamma$ and $z$, but not the dependence of the density of states on $z$ nor the equation of state. We can nevertheless obtain a similar free energy which incorporates equation \ref{gamz3}:
\begin{equation} F(\gamma,z)=\alpha \gamma - 2\ln \gamma - c_{h} t^{3} \left[\ln\left(\frac{\gamma}{B_{h}t^{3}}\right)+1\right]\label{free_empirical_hertz} \end{equation}
This then makes a prediction for the density of states and the equation of state for a system with hertzian interactions:
\begin{equation} \Omega(\gamma,z) \sim \gamma^{2+c_{h}t^{3}} \quad \text{and} \quad \alpha = \frac{2}{\langle \gamma \rangle}(2+c_{h}\langle t\rangle^{3}) \end{equation}
More generally, for systems with a contact interaction of the type used in \cite{epitome} with a power $\delta$, we predict an effective free energy
\begin{equation} F(\gamma,z)=\alpha \gamma - 2\ln \gamma - c t^{2(\delta-1)} \left[\ln\left(\frac{\gamma}{B t^{2(\delta-1)}}\right)+1\right] \end{equation}

\subsection{Phase transition in the mean-field theory}
We now investigate the jamming transition in the context of the free energy equation \ref{free_empirical_spring}, at first for a system with harmonic interactions. The constants in the free energy can be scaled out to give
\begin{equation} F(\gamma,x)=\alpha \gamma - 2\ln \gamma - x^{2} \left[\ln\left(\frac{\gamma}{x^{2}}\right)+1\right],\label{free_scaled} \end{equation}
with the scaled variables $x \equiv c^{1/2}(z-z_{iso})$, $\gamma \equiv \frac{c}{B}\gamma$ and $\alpha \equiv \frac{B}{c}\alpha$.

Minimizing equation \ref{free_scaled} with respect to its fields $\gamma$ and $x$ allows us to extract the scaling of the ensemble averages $\langle \gamma \rangle$ and $\langle x \rangle$ with the inverse temperature $\alpha$. We find 
\begin{equation} \langle \gamma \rangle = \frac{2}{\alpha-1} \quad \text{and} \quad \langle x \rangle = \left(\frac{2}{\alpha-1} \right)^{1/2} \label{gam_z_alpha_mf}, \end{equation}
which is consistent with the equation of state at $z_{iso}$ and the relation between $\gamma$ and $z$, as expected.

\begin{figure}[h]
\begin{center}
\includegraphics[width = 0.49\columnwidth, trim = 0mm 0mm 10mm 0mm, clip]{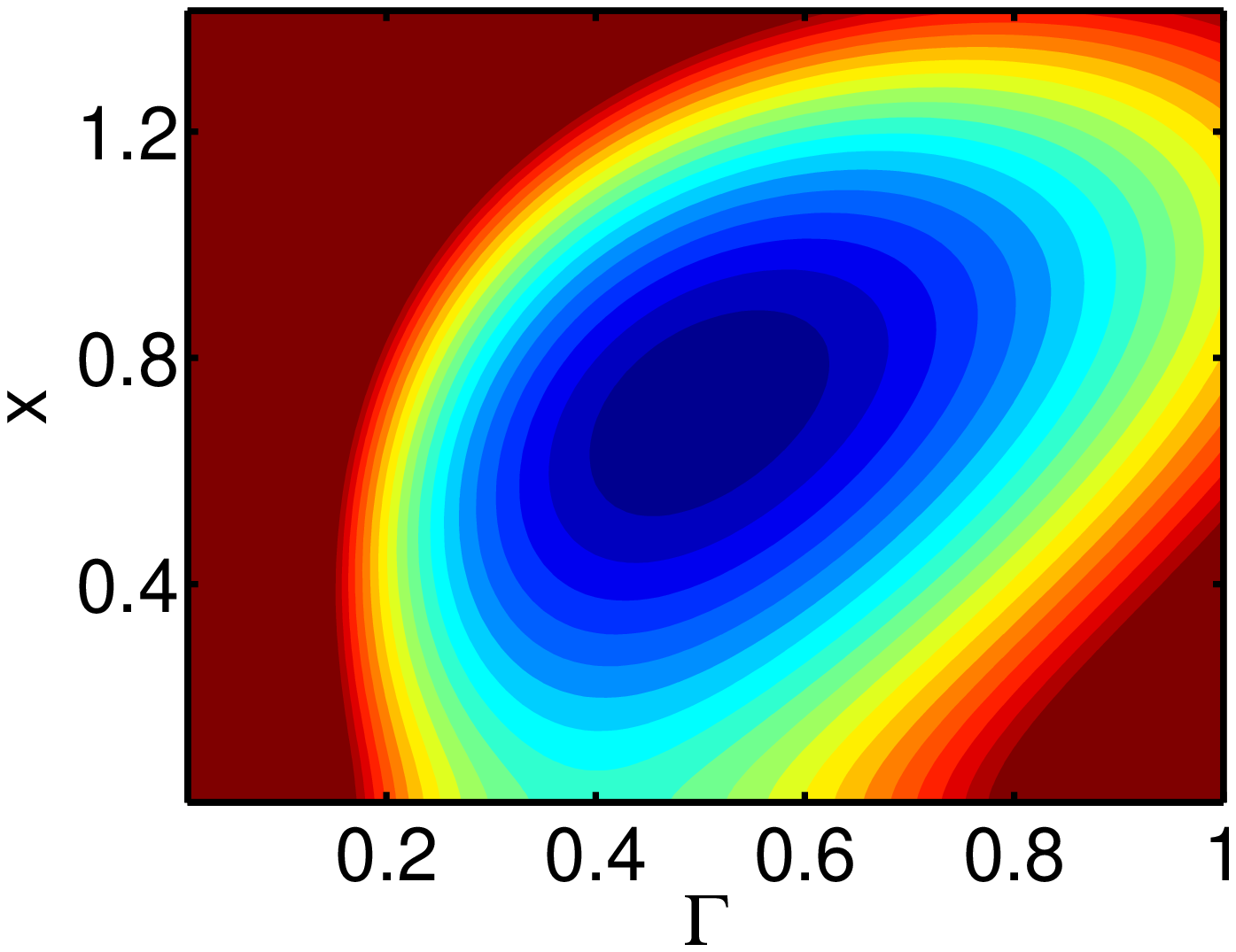}
\includegraphics[width = 0.49\columnwidth, trim = 0mm 0mm 10mm 0mm, clip]{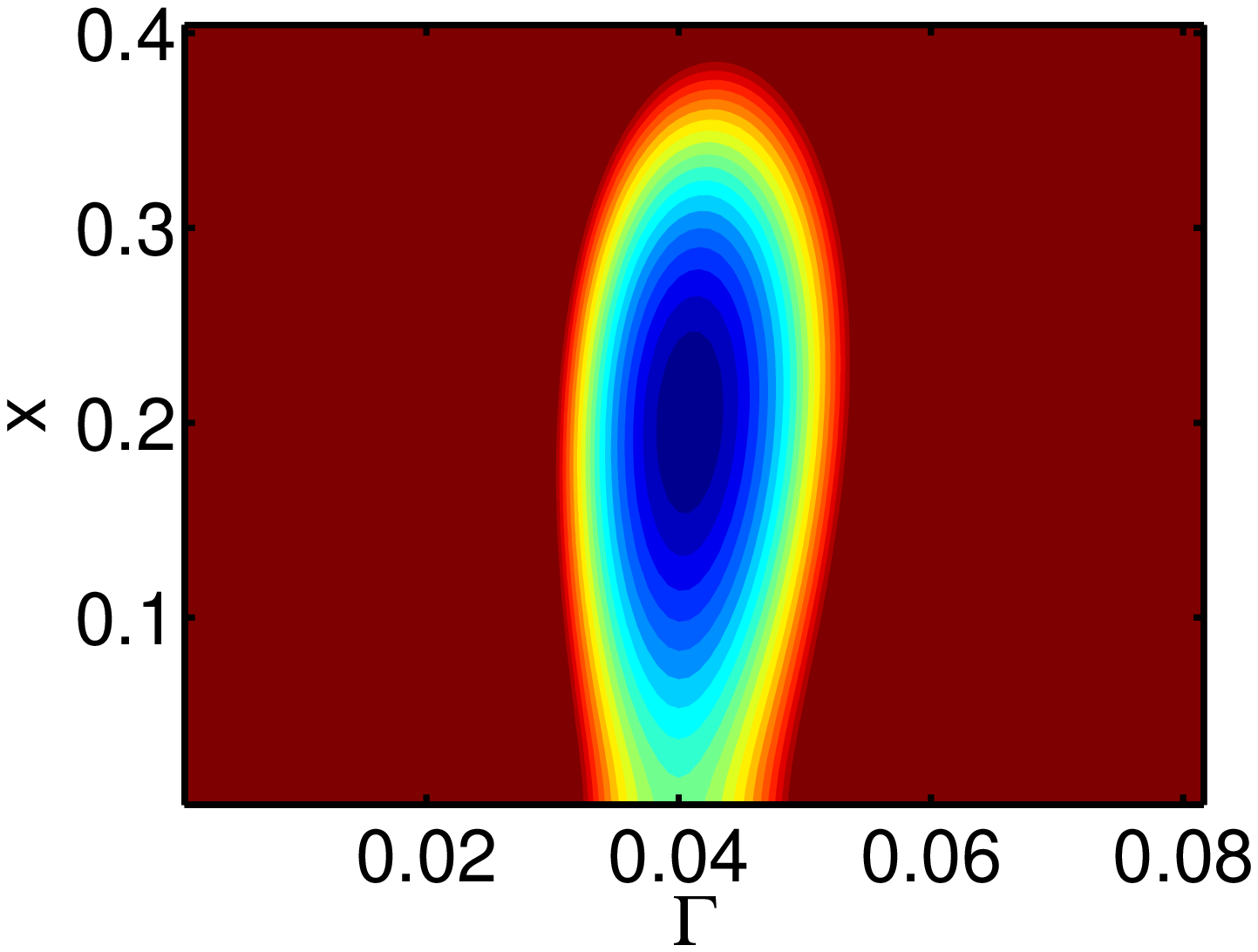}
\end{center}
\caption{Contour plot of the free energy equation \ref{free_scaled} for $\alpha=5$ (left) and for $\alpha=50$ (right), approaching the transition. The scale for the contours in both cases is the height of the free energy barrier to $x=0$, $F(\langle \gamma\rangle,\langle x \rangle) - F(\langle \gamma \rangle, 0)$. \label{free_xgam}}
\end{figure}

\paragraph*{Form of the transition}
The jamming transition is the point $\langle \gamma \rangle \rightarrow 0$ and $\langle x \rangle \rightarrow 0$, i.e. $\alpha \rightarrow \infty$. Figure \ref{free_xgam} shows the change in shape of the free energy approaching the jamming transition, and we observe a narrowing of the width of $F$ around the minimum in the $\gamma$-direction, while the width in the $x$-direction increases.

The signature of a second-order phase transition is a divergence of the fluctuations of the order parameters $x$ or $\gamma$ as the transition is approached, i.e. a vanishing curvature of the free energy at its minimum at the transition \cite{chandlerbook}. We calculate the Hessian matrix at the minimum of $F$ and we obtain
that in the limit $\alpha \rightarrow \infty$, the eigenvectors of this matrix become the $\gamma$- and $x$-directions, with eigenvalues that scale as $\lambda_{\gamma} \sim \alpha^{2}$ and $\lambda_{x} =4$. This shows that the minimum in the $\gamma$-direction narrows drastically as the transition is approached, consistent with the well-definened equation of state we observe, and with the divergence of the stiffness in the field theoretical picture. The magnitude of the fluctuations in $x$, however, is \emph{independent} of the distance from the jamming transition, and does not diverge.

With $\gamma$ and $x$ as order parameters, the jamming transistion does not have the properties of a second-order phase transition. Neither can it be described as a first-order phase transition since it lacks a second minimum of the free energy. This reflects the fact that within the stress-ensemble framework, a granular packing can only be in a single state, i.e. jammed. Unstable configurations cannot be accounted for within the framework.

\subsection{A new order parameter \label{sec:gam_z2}}
\paragraph*{Motivation}

The work of Wyart et al. \cite{Wyart,Witten_DOS} and Ellenbroek et al. \cite{Ellenbroek} shows that there is an excess of low-energy deformation modes in soft granular packings close to the jamming transition. The authors link the soft modes to very low energy rearrangements in which the grains slip past each other. They also show that the scale of the rearrangements is determined by the distance of the packing from the isostatic point. A simplified version of the argument considers a packing of size $l$ at mean contact number $z$, so that the number of force components in excess of the force and torque balance requirements in the packing is given by $\delta n_{f} = N\delta z/2 \sim l^{d} \delta z$ ($\delta z=z\!-\!z_{iso}$). If we cut the boundary of the system, we remove of the order $l^{d-1}$ force components, and create an unstable region if $l^{d-1} \geq l^{d} \delta z$.  This predicts a characteristic length scale $l^{*} \sim \delta z^{-1}$ below which the packing is locally unstable, and this length scale diverges as the jamming transition is approached. The scale of the soft modes is set by $l^{*}$, so that their energy becomes $\omega_{s}^{2} \sim (l^{*})^{-2} \sim (\delta z)^{2}$. 

The energy for the soft modes for a packing at pressure $p=\gamma/A$ diminishes proportionally to the stress the soft modes cause at finite compression, which is proportional to $p$ for a harmonic interaction potential. Therefore, the frequency of the soft modes at positive pressure scales as $\omega_{s}^{2} \sim A (\delta z)^{2} - \gamma$, where $A$ is a constant.

Every mechanically stable packing must satisfy the inequality 

\begin{equation} 0<\delta E \leq A (z\!-\!z_{iso})^{2} - \gamma, \label{gam_z_inequality}\end{equation}
where $\delta E$ is the energy of the lowest-energy displacement eigenmode of the system.

\begin{figure}
\begin{center}
\includegraphics[width = 0.7\columnwidth, trim = 15mm 150mm 20mm 10mm, clip]{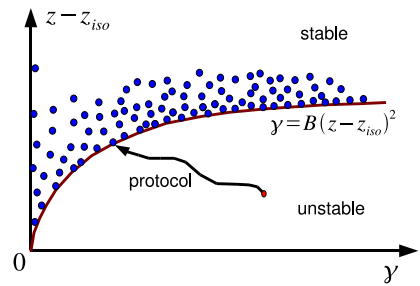}
\end{center}
\caption{Phase diagram derived from equation \ref{gam_z_inequality}. The red line marks the phase boundary between the stable and unstable packings and the blue dots show stable packings. The sampling effect of the conjugate gradient minimization protocol is illustrated by the black arrow leading from the initial configuration to the final stable packing on the phase boundary.}
\label{phase_diag_gamz}
\end{figure}

The equation predicts a phase diagram for static granular packings, with stable packings above the $\gamma=A(z\!-\!z_{iso})^{2}$-line, and unstable packings below it (see Figure \ref{phase_diag_gamz}).

%% Silke 27.08: Need to mention marginally stable in a correct and clear way

The simulated packings always lie \emph{on} the boundary between the stable ($\gamma < B (z\!-\!z_{iso})^{2}$) and unstable ($\gamma > B (z\!-\!z_{iso})^{2}$ regimes, i.e they are \emph{marginally stable}. Wyart et al. have suggested a mechanism to explain this \cite{Wyart}: The conjuguate gradient minimization will decrease the total potential energy of the system by reducing the overlap between initially random disks just enough to reach mechanical equilibrium. During the procedure, on average, $\gamma$ will reduce with the energy, while the contact number $z$ will increase. Stopping the process at the first stable configuration reached biases the process towards configurations with high $\gamma$ and low $z$ compared to a flat sample of all the stable configurations. The same bias is not immediately encountered in other methods used in the literature to create mechanically stable configurations, such as tapping \cite{Jaeger_Nagel_compaction} or shearing \cite{Behringer_logshear}. More generally, the marginally stable configurations should be encountered in any protocol which does not allow the system to thermalize, i.e. explore the phase space of jammed configurations, but instead does an infinitely rapid quench to the nearest stable packing.

Figure \ref{z2gamma} shows that as we approach the isostatic point, the relative fluctuations around the stability line derived above increase. Therefore, it is natural to investigate the variable $u=\frac{x^{2}}{\gamma}$ which takes the value of $1$ on the stability line, and $u>1$ in the stable region.

\begin{figure}[h]
\begin{center}
\includegraphics[width = 0.49\columnwidth, trim = 0mm 0mm 10mm 0mm, clip]{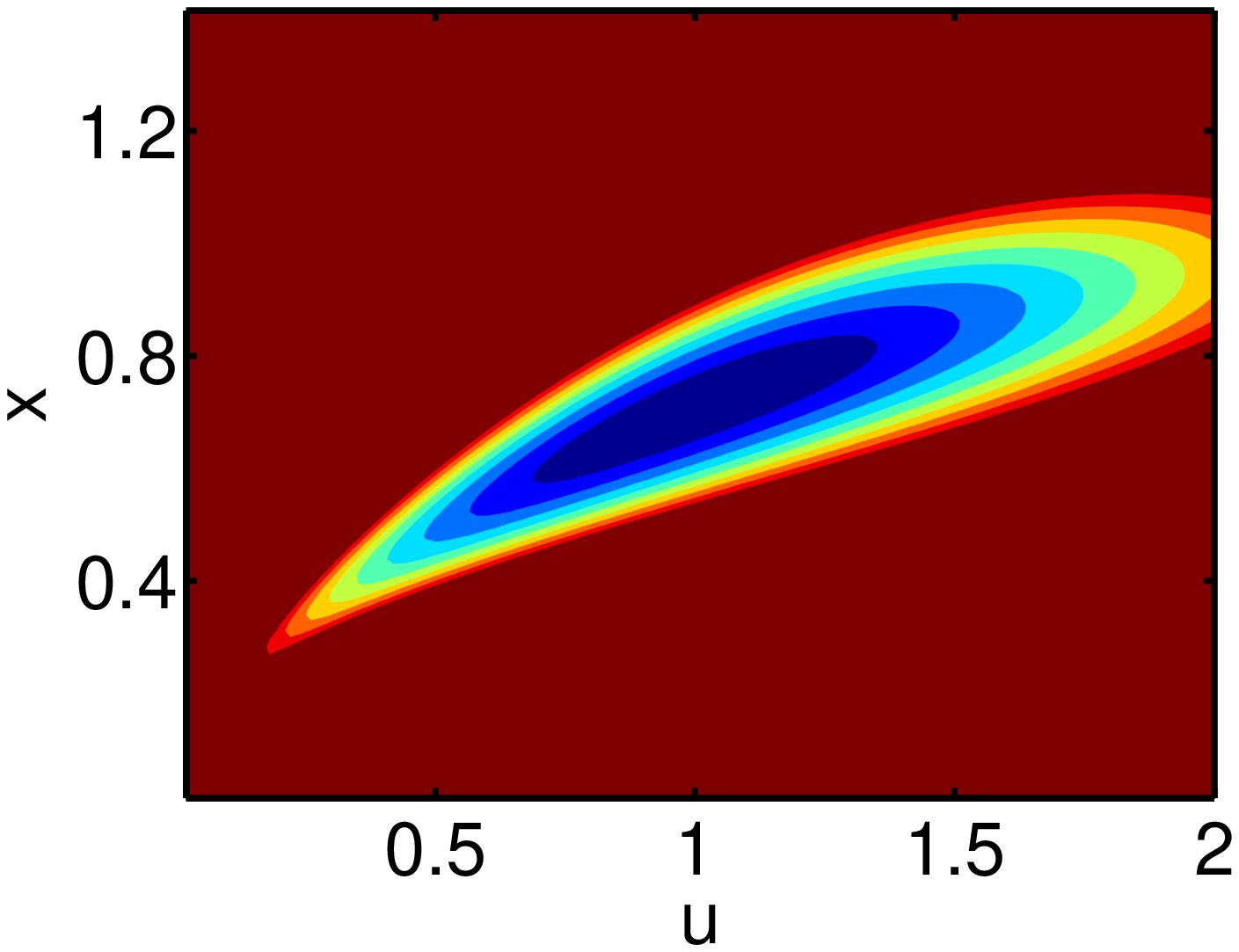}
\includegraphics[width = 0.49\columnwidth, trim = 0mm 0mm 10mm 0mm, clip]{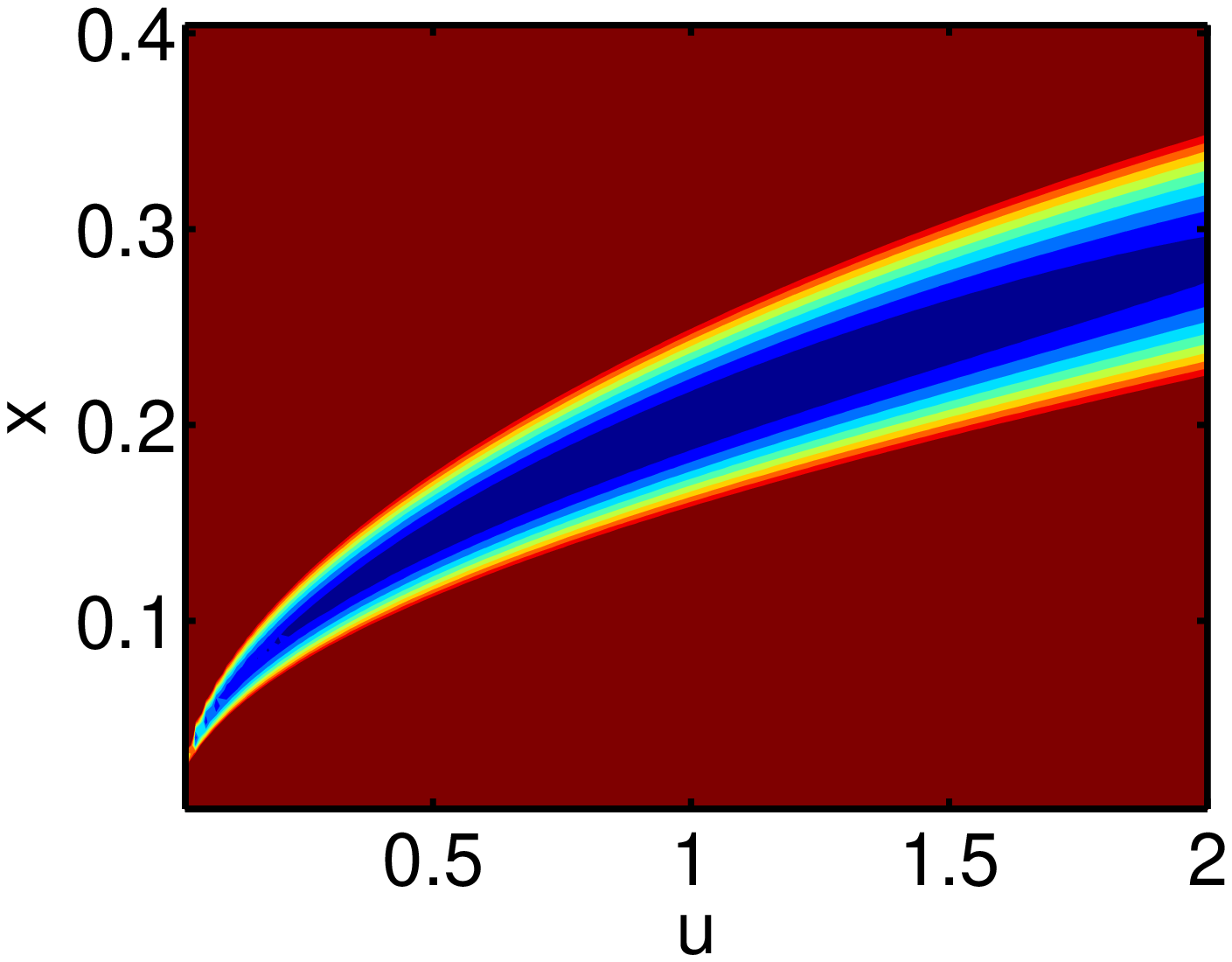}
\end{center}
\caption{Contour plot of the free energy equation \ref{free_ux} for $\alpha=5$ (left) and for $\alpha=50$ (right), approaching the transition. The scale for the contours in both cases is the value of the second derivative of $F$ with respect to $u$ at the minimum of $F$. \label{Fplot_ux}}
\end{figure}

\paragraph*{Mean field theory in $u$-$x$ coordinates}
We can rewrite the free energy equation \ref{free_scaled} in terms of the new variables $u$ and $x$
\begin{equation} F(u,x)=\alpha \frac{x^{2}}{u} - 2\ln \left[\frac{x^{2}}{u}\right] - x^{2} \left(1-\ln u\right),\label{free_ux} \end{equation}
and investigate the properties of the transition in function of $u$ and $x$. The position of the minimum is at $\langle u\rangle=1$ and $\langle x \rangle = \left(\frac{2}{\alpha-1}\right)^{1/2}$, which is compatible with the results obtained in the $(\gamma,x)$-coordinates. Figure \ref{Fplot_ux} shows the evolution of the shape of $F$ as the jamming transition is approached.
We observe a striking asymmetry develop in the width of the minimum of $F$ as the jamming transition is approached. It is clear that there exists a direction along which the second derivative of $F$ vanishes, and hence the fluctuations diverge. The hessian matrix is given by
\begin{equation} \frac{\partial^{2} F}{\partial u,\partial x} =
\left(\!\!
\begin{array}{cc}
  2 \frac{\alpha}{\alpha-1} & - 2\left[2(\alpha-1)\right]^{1/2} \\
 -2 \left[2(\alpha-1)\right]^{1/2}  & 4(\alpha-1) \\
\end{array}
\!\!\right).
\end{equation} 

The eigenvalues close to unjamming, for large $\alpha$, become $\lambda_{-} \simeq \frac{1}{\alpha}$ and $\lambda_{+} \simeq 2\alpha$.
Hence the curvature of the free energy vanishes along the direction of $\vec{v}_{-}$, whose angle with the $u$-axis is just $\tan(\theta)=v_{-}^{2}/v_{-}^{1}$. At the unjamming transition $\alpha
\rightarrow \infty$, this angle vanishes as $\tan(\theta)\simeq (2\alpha)^{-1/2}$, and the flat eigendirection becomes the $u$-axis.

So in the limit $\alpha\rightarrow\infty$, we observe a diverging susceptibility in the $u$-direction, that is the fluctuations around the stability line $\gamma=x^2$ diverge. This indicates critical behavior at the jamming transition, further evidence that Point J is a critical point if approached along the $T=0$-line.

\begin{figure}[ht]
\begin{center}
\includegraphics[width = 0.9\columnwidth, trim = 0mm 0mm 10mm 0mm, clip]{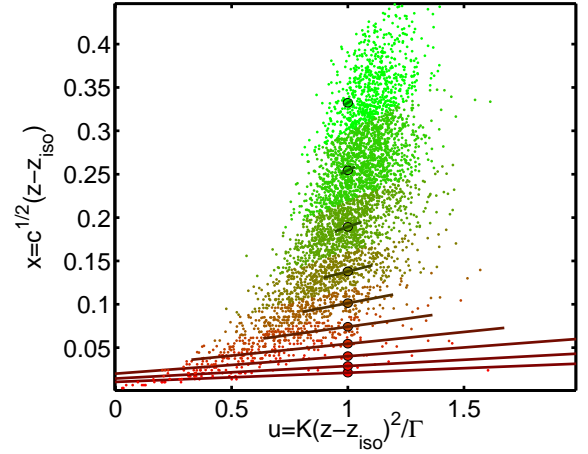}
\end{center}
\caption{Data points from the $N=1024$-configurations with harmonic interactions grouped by different values of $\alpha$ (red: high $\alpha$, green: low $\alpha$) plotted in the $x-u$ diagram. For each $\alpha$, the circle of the same color marks the minimum of $F$ at that $\alpha$. The line through each circle is in the direction of the eigenvector along which the susceptibility diverges. Its length is proportional to $1/\lambda_{-}$, i.e. proportional to the expeted fluctuations.}
 \label{u_data_compare}
\end{figure}

\paragraph*{Comparison to simulation data}
The first prediction of the field theory in $u-x$-coordinates is that all the $(u,x)$ data points associated to individual configurations should cluster around $u=1$, after an appropriate rescaling. Second, we predict that the $x$ value of of the data points for a given $\alpha$ clusters around $\langle x \rangle = \left(\frac{2}{\alpha-1} \right)^{1/2}$. Figure \ref{u_data_compare} shows configurations from throughout the jammed region, grouped by their values of $\alpha$ (indicated by the color of the data points). The data cluster around $u=1$, with large fluctuations of for data points at higer $\alpha$ (in red). The circles, which are in the same color as the data points mark the minimum of $F$ for the $\alpha$ associated to that color.  The line through each circle is in the direction of the eigenvector along which the susceptibility diverges. Its length is proportional to $1/\lambda_{-}$, i.e. proportional to the magnitude of the expected fluctuations. We observe that the data points do indeed cluster around the minimum and follow the direction given by the eigenvector. In the limit of the jamming transition, the spread of the data points at a given $u$ becomes very large, and parallel to the $u$-direction, as expected.
\begin{figure}[ht]
\begin{center}
\includegraphics[width = 0.9\columnwidth, trim = 0mm 0mm 10mm 0mm, clip]{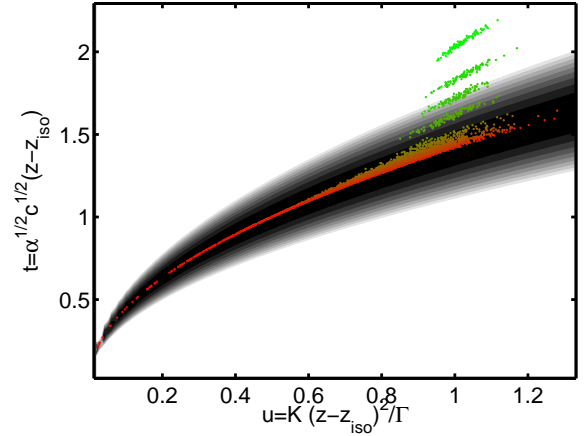}
\end{center}
\caption{Same as Figure \ref{u_data_compare} after the $x$-axis has been rescaled by $\alpha^{1/2}$, superimposed on a contour plot of the scaling limit of the free energy equation \ref{free_scallim} (see text).}
 \label{u_data_scaling}
\end{figure}

We can obtain a scaling form of the free energy close to the jamming transition. The $x^{2}(1-\ln u)$ term in equation \ref{free_ux} becomes vanishingly small compared to the other terms in this limit, and if we define $t=\alpha^{1/2} x$ we can write the free energy (up to a constant) as
\begin{equation} F(t,u) = \frac{t^2}{u}-2\ln\left(\frac{t^{2}}{u}\right) \label{free_scallim}.\end{equation} 
Figure \ref{u_data_scaling} shows a contour plot of $F(t,u)$, and superimposed on it the same data points as in Figure \ref{u_data_compare}, multiplied by their individual $\alpha^{1/2}$. We obtain an excellent collapse of the data for all the configurations near Point J, and the divergent fluctuations along the minimum of $F$ can clearly be seen.

%*************************************************************************************************************************

\subsection{Link of the mean field theory to the field theory}
For the field theory derived in section 3 and the mean field theory explored above to be consistent with each oher, the partition function has to take the following form:
\begin{align}
&Z_{m}(\alpha) = \int d\gamma dz \exp\left( -m F(\gamma,x)\right) \nonumber\\
	 &= \int d\gamma dz \int D\psi \exp\left[-\frac{K_{0}}{\gamma^{2}} \int \frac{d^{2}q}{2\pi^{2}} \left( (2+c(z-z_{iso})^{2})q^{4} \right.\right. \nonumber \\
	&\left.\left.+\xi_{2}^{2}q^{6}+\xi_{4}^{4}q^{8}\right)|\psi_{q}|^{2}\right] \exp(-\alpha\gamma)
\end{align}
We can show this link in two ways. The simpler method is to expand the mean field theory to second order about its minimum in $\gamma(\vec{r}) = \gamma +\nabla^{2}\psi(\vec{r})$, while keeping the scaled contact number $x$ as a parameter. We obtain
\begin{equation} F(\psi,x) = F(\gamma,x)-\int d^{2} r \frac{2+c x^{2}}{\gamma^{2}} |\nabla^{2} \psi|^{2}, \end{equation}
which is nothing but the stiffness part of the field theory. This shows consistency since the purely microscopic length scales $\xi_{2}$ and $\xi_{4}$ are not expected to contribute to the mean field free energy. 

Alternatively, we can perform the path integral in $\psi$ and see if we obtain the entropic part of the mean field free energy from the logarithm of the microcanonical partition function. Although this can be done exactly, since we limited ourselves to gaussian terms, there are problems with this approach. The field theory is an expansion in powers of $\psi$, and we have kept only the lowest order. This is sufficient to calculate the \emph{moments} of $\psi$, like the structure factor, but it is probably inadequate to accurately calculate the generating functional $Z(\gamma,x)$. A proper treatment of the problem would need knowledge of the full power series and then a sophisticated renormalization approach, for a two-dimensional problem.

\begin{figure}[h]
\begin{center}
\includegraphics[width = 0.8\columnwidth, trim = 0mm 0mm 10mm 5mm, clip]{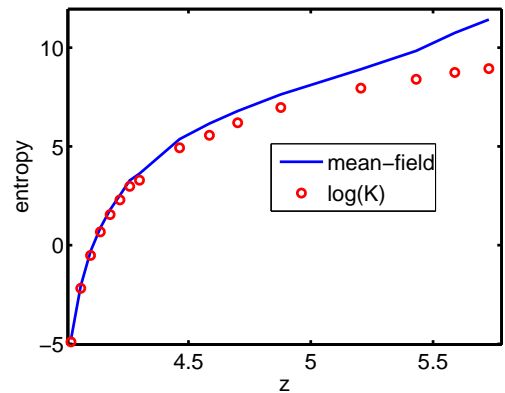}
\end{center}
\caption{Comparison between the mean-field theory (blue line) and the field theoretical result for $S(\gamma,z)$, using the values of $(\gamma,z)$ from the simulations for the latter (red dots). }
\label{compare}
\end{figure}

Nevertheless, some progress can be made and we can test the consistency between the two forms to lowest order in $\gamma$ and we obtain\cite{thesis}

\begin{equation}
S(\gamma,z) =S_{0} + N\left(2\ln\gamma-\ln(2+\tilde{c}(z-z_{iso})^{2})\right). \end{equation}
Comparing this to the mean-field result, we find that the isostatic limit of both expressions, $2\ln\gamma$, is identical. We recover part of the $z$-dependence if we expand the logarithm of the contact number $\ln(2+\tilde{c}(z-z_{iso})^{2})\approx \ln2+1/2 \tilde{c}(z-z_{iso})^{2}$. The field theory does surprisingly well in a numerical comparison to the mean-field result. Figure \ref{compare} shows both for the pairs of $(\gamma,z)$ corresponding to the structure factor plotted in Figure \ref{struplot}. This is likely due to the fact that the $(\gamma,z)$-pairs investigated are all quite close to the line of $u=1$, so that the logarithmic term $\ln\left(\frac{\gamma}{B (z-z_{iso})^{2}}\right)$ vanishes. On the line $u=1$, both expressions reduce at first order to
\begin{equation} S=S_{0} + N\left[ 2\ln\gamma -c x^{2}\right], \end{equation}
with a value $c=2.8\pm0.5$ from the mean field theory and a value $c=\tilde{c}/2 = 2\pm 0.25$ from the field theory.

\section{Conclusions}

\subsection{Summary}
We have investigated the properties of jammed granular assemblies, approaching Point J and within the jammed region, using a newly formulated stress ensemble. The jamming transition can be analyzed within this framework.

The force moment tensor of a system can be written as a boundary term, which makes it a conserved quantity under internal rearrangements. We use this conservation law to define a canonical ensemble in section \ref{chap:ensemble}. The conjuguate variable to the force moment tensor defines then a granular analog to temperature, that is closely related to the Edwards definition of angoricity.

We review our previous test \cite{prl07} of the ensemble, and obtain an equation of state for the granular temperature within the jammed region. 
The form of the equation of state is a universal property at the isostatic point, a conjecture that we confirm through an exact calculation, though we also detect evidence for short-range correlations. 

In section 3, using the constraint-free Airy stress function, we build a field theoretical model for the jammed range based on symmetry arguments and the analysis of the pressure fluctuations in the simulated configurations. The jamming transition appears as a divergence of the stiffness of the leading term, so that the entropy of jammed packings tends to zero at the transition.

Finally, in section 4 we combine all the relations obtained from simulation into a phenomenological mean-field theory. We investigate the jamming transition again to determine the correct order parameter. Inspired by a stability argument proposed by \cite{Wyart}, we use an order parameter which measures the deviatins from the stability line linking mean pressure and mean number of contacts. The mean field theory then predicts divergent fluctuations in this order parameter at Point J, lending weight to the interpretation of J as a critical point.

\subsection{Limitations of the ensemble, predictions and further tests}
\paragraph*{Limitations and applicability}
The stress ensemble, with its tensorial conserved quantity and tensorial temperature, can be applied to a wide range of systems. The only conditions are that the system has to be in mechanical equilibrium, and that there can be no long-range interactions between the particles. This excludes any kind of driven system (exceptions below), or system with a temperature, but it includes systems with attractions, like colloids, or packings with friction. Anisotropic systems, either made of nonspherical particles and/or with an imposed shear in addition to a hydrostatic pressure, are a major area where we expect the stress ensemble to be useful.

The analysis of the simulated packings, the field theory, the mean field theory and the interpretation of Point J in this context are limited to isotropic, frictionless packings of round grains. Lifting any one of these restrictions fundamentally changes the nature of the system. 

For packings with an imposed shear, we expect the ``Boltzmann factor'' $\exp(-\text{Tr}(\hat{\alpha}\hat{\Sigma}))$ to reduce to a factor $\exp(-\alpha_{p} p)$ featuring the presssure $p=1/2 (\sigma_{1}+\sigma_{2})$ and a pressure-temperature $\alpha_{p}$ and a factor $\exp(-\alpha_{s} \tau)$ featuring the shear $\tau=1/2 (\sigma_{1}-\sigma_{2})$ and a shear-temperature $\alpha_{s}$. Here $\sigma_{1}$ and $\sigma_{2}$ are the principal stresses for the global stress tensor. We have extended the field theory presented here to the pure shear case, and are in the process of testing its predictions against simulations and experiments\cite{Field_Shear_draft}.

The introduction of friction, or of anisotropic grains transform the isostatic point into a broader region within which marginally stable packings occur. It is then doubtful if Point J is a single, well defined singular point for these systems, and it will be interesting to investigate the field theoretic framework for these cases.

\subsection{Predictions}
In principle, all the relations obtained from the application of the ensemble to the simulated system are predictions which can be verified in other simulations or in experimental systems which are close to frictionless, round particles, like bubble rafts. There are however several major caveats. 

First, the density of states at larger compressions is dependent on the method of preparation of the sample, as discussed in section 4 and it remains to be seen which range of (simulation or experimental) protocols selects for marginally stable packings. The behavior close to the jamming transition should not be affected, though. 

Second, most of the predictions were made in the $\alpha$-canonical ensemble which is difficult to reproduce in experiments or in simulations since there are no means of imposing $\alpha$ externally, as is normally done with the temperature in thermal systems. It is feasible, though, to work in the \emph{microcanonical} ensemble where the external stress (or the hydrostatic pressure, for isotropic systems) is fixed. The predictions of the canonical ensemble can be adapted to the microcanonical ensemble by carrying out a Legendre transform if the subsystem investigated is sufficiently large (i.e. $m>>1$). For smaller subsystems, especially for the force distribution and the single-grain pressure distribution, the local correlations make the thermodynamic approximation break down.

The best prediction is the equation of state $\alpha = \frac{z_{iso} N}{2 \Gamma}$, together with the coarse-grained density of states $\Omega(\Gamma_{m})\sim \Gamma^{2m}$ close to Point J for $2d$ frictionless round grains.  It is a universal property which we expect to hold regardless of preparation method and the choice of the ensemble.

\subsection{Further tests}
There are several interesting methods by which the stress ensemble can be explored.

When a hot and a cold body come into touch, the second law of thermodynamics dictates that heat will flow from the hot body to the cold body until both of them arrive at the same temperature, regardless of the composition of the two bodies. In the context of the stress ensemble this means that two compartments at different $\alpha_{1} \neq \alpha_{2}$, if brought into touch, will transmit stress from the compartment with lower $\alpha$ (i.e. higher granular temperature) to the compartment with the higher $\alpha$, until $\alpha_{1}=\alpha_{2}$. To set up a numerical experiment testing this situation, one could prepare two sets of simulated configurations similar to the ones tested in \cite{prl07}, determine their respective equations of state, and then bring them into touch. Then one can measure the $\Gamma_{m}$-distributions in the two halves, determine their respective $\alpha$ via the equations of state and test if both values of $\alpha$ match.

The only dynamical situation where the stress ensemble can be applied is a quasistatic motion where the system evolves through a sequence of equilibrium states. One example of this is a system under quasistatic shear, slow enough for it to be characterized by a sequence of stress buildups and eventual rearrangements. It is in this regime that we expect that the $SGR$ formalism  \cite{Sollich} can be adaped from the Boltzmann ensemble to the stress ensemble, as we have done for a toy model\cite{bob_toymodel}.

Another situation of the same type is a system which is periodically tapped so that it rearranges itself into a new configuration. In contrast to the same setup in the context of the Edwards ensemble, we need a system with the same externally imposed stress after each rearrangement. For this system, or for the quasistatically sheared system, we can then employ a granular version of the fluctuation-dissipation theorem (FDT), as has been done by Song et al. for the Edwards ensemble \cite{Makse_tracer}. Let $F$ be an external perturbing force, and $x(t)$ be the position of a tracer particle. Then the FDT links the fluctuations in the particle position to its mean response to the perturbation:
\begin{equation}  \langle (x(t)-x(0))^{2} \rangle = \frac{2}{\alpha} \frac{\langle x(t)-x(0)\rangle}{F} \end{equation}
The driving force, provided by gravity in the work of Song et al., should be replaced by e.g. the magnetic force on a metallic tracer particle since the stress ensemble is sensitive to a gravitational field. The values of $\alpha$ extracted by this method should then only depend on the external stress, not on the type of material employed or the magnitude of the driving force.

\subsection{Outlook}
This work underscores that the experimental conditions under which a granular material is examined are crucially important. The framework we have introduced permits us to distinguish between systems in a canonical stress ensemble or in a microcanonical stress ensemble, and between different stress states. The statistical framework can be used to predict correlations functions of the stress under these different conditions.

The jamming phase diagram masks a more complicated reality, arising from the sensitivity of the jamming transition to the prepared state of the packing (see \cite{Bulbul_Bob} for a review).  The stress ensemble provides a concrete framework for understanding  the behavior of granular materials at the jamming transition and within the jammed region.

\begin{acknowledgments}
We acknowledge many useful discussions with Bob Behringer, Corey O'Hern, Olivier Dauchot and Brian Tighe.  This work was supported by NSF-DMR 0549762.  BC acknowledges the hospitality of the Aspen Center for Physics, where some of this work was done.
\end{acknowledgments}

\bibliography{thesis_references}

\end{document}